\documentclass[preprint]{iucr}              
     \journalcode{S}              
\papertype{FA}
\usepackage{amsmath,amsfonts,amssymb}
\usepackage{graphicx}
\usepackage{textcomp}
\usepackage{multirow}
\usepackage{float}
\newcommand{\vf}{\boldsymbol{f}}
\newcommand{\vp}{\boldsymbol{p}}
\newcommand{\vh}{\boldsymbol{h}}

\newcommand{\vfUNet}{\vf_\text{U-Net}}
\newcommand{\vfUNetP}{\vf_\text{U-Net, PWLS}}
\usepackage{color}

\begin{document}                  



\title{Limited Angle Tomography for Transmission X-Ray Microscopy Using Deep Learning}
\shorttitle{Limited Angle Tomography Using Deep Learning}


\cauthor[a]{Yixing}{Huang}{yixing.yh.huang@fau.de}{}
\cauthor[b,c]{Shengxiang}{Wang}{wangsx@ihep.ac.cn}{}
\author[d]{Yong}{Guan}{}{}
\author[a,e]{Andreas}{Maier}{}{}

\aff[a]{Pattern Recognition Lab, Friedrich-Alexander-Universit\"at Erlangen-N\"urnberg, Erlangen 91058, \country{Germany}}
\aff[b]{Spallation Neutron Source Science Center, Dongguan, Guangdong 523803, \country{China}}
\aff[c]{Institute of High Energy Physics, Chinese Academy of Sciences, Beijing 100049, \country{China}}
\aff[d]{National Synchrotron Radiation Laboratory, University of Science and Technology of China, Hefei, Anhui 230026, \country{China}}
\aff[e]{Erlangen Graduate School in Advanced Optical Technologies (SAOT),
Erlangen 91052, \country{Germany}}


\shortauthor{Yixing Huang \textit{et al}.}






\maketitle                        

\begin{synopsis}
A deep learning method for limited angle tomography in synchrotron radiation transmission X-ray microscopies and a demonstration of its application in 3-D visualization of a chlorella cell.
\end{synopsis}

\begin{abstract}
In transmission X-ray microscopy (TXM) systems, the rotation of a scanned sample might be restricted to a limited angular range to avoid collision to other system parts or high attenuation at certain tilting angles. Image reconstruction from such limited angle data suffers from artifacts due to missing data. In this work, deep learning is applied to limited angle reconstruction in TXMs for the first time. With the challenge to obtain sufficient real data for training, training a deep neural network from synthetic data is investigated. Particularly, the U-Net, the state-of-the-art neural network in biomedical imaging, is trained from synthetic ellipsoid data and multi-category data to reduce artifacts in filtered back-projection (FBP) reconstruction images. 
The proposed method is evaluated on synthetic data and real scanned chlorella data in $100^\circ$ limited angle tomography. 
For synthetic test data, the U-Net significantly reduces root-mean-square error (RMSE) from $2.55 \times 10^{-3}$\,\textmu m$^{-1}$ in the FBP reconstruction to $1.21 \times 10^{-3}$\,\textmu m$^{-1}$ in the U-Net reconstruction, and also improves structural similarity (SSIM) index from 0.625 to 0.920. With penalized weighted least square denoising of measured projections, the RMSE and SSIM are further improved to $1.16 \times 10^{-3}$\,\textmu m$^{-1}$ and 0.932, respectively.
For real test data, the proposed method remarkably improves the 3-D visualization of the subcellular structures in the chlorella cell, which indicates its important value for nano-scale imaging in biology, nanoscience and materials science.
\end{abstract}


\section{Introduction}

Transmission X-ray microscopy (TXM) has become a very powerful technology for nano-scale imaging in various fields \cite{wang2000soft,chao2005soft,sakdinawat2010nanoscale,wang2016nanotechnology}, including materials science \cite{andrews2011transmission,nelson2012operando}, chemistry \cite{de2008nanoscale,wang2015use}, and biology \cite{shapiro2005biological,wang20153d}. With projection images acquired in a series of rotational angles, tomographic images can be reconstructed according to computed tomography (CT) technologies for 3-D visualization of scanned samples. In such applications, TXM is also called X-ray nano-CT \cite{shearing2011using,brisard2012morphological,liu2018quantitative}. A TXM system typically consists of a central stop, a condenser, a sample holder, an objective zone plate, and a CCD detector, with X-rays generated from synchrotron radition or a high-end X-ray source. TXMs typically utilize a pin as the sample holder \cite{holler2017omny}, e.\,g., tip versions for pillar samples, glass capillaries for powder samples, copper capillaries for high pressure cryogenic samples, and grids for flat samples. For tips and capillaries, rotating a sample in an sufficient angular range is not a problem. However, for grids, collision between the grid and the zone plate, which is very near to the rotation axis in TXM systems, might happen in large scan angles. In addition, for flat samples, the lengths of X-rays through the sample increase rapidly at high tilting angles \cite{barnard1992360,koster1997perspectives}, which introduces a high level of scattering and reduces image contrast. Therefore, in these situations, the problem of limited angle tomography arises. 

Limited angle tomography is a severely ill-posed inverse problem \cite{davison1983ill,louis1986incomplete,natterer1986mathematics,quinto2006introduction}. Using microlocal analysis, edges that are tangent to available X-rays can be well reconstructed while those whose singularities are not perpendicular to any X-ray lines cannot be reconstructed stably \cite{quinto1993singularities,quinto2006introduction}. So far, many algorithms have been developed to deal with this task. Among these algorithms, extrapolating missing data is the most straightforward way for limited angle tomography. The iterative Gerchberg-Papoulis extrapolation algorithm \cite{gerchberg1974super,papoulis1975new} based on band-limitation properties of imaged objects has been demonstrated beneficial for improving image quality of limited angle tomography \cite{defrise1983regularized,Qu2008An,Qu2009Landweber,Huang2018Papoulis}. In addition, data consistency conditions, e.\,g. the Helgason-Ludwig consistency conditions \cite{helgason1965radon,ludwig1966radon}, provide redundancy and constraint information of projection data, which effectively improves the quality of extrapolation \cite{louis1980picture,louis1981approximation,willsky1990constrained,kudo1991sinogram,huang2017Restoration}. Nevertheless, such extrapolation methods have only achieved limited performance on real data, which typically contain complex structures and are very difficult to extrapolate.

Iterative reconstruction using sparse regularization technologies, particularly total variation (TV), has been widely applied to image reconstruction from insufficient data. TV methods employ the sparsity information of image gradients as a regularization term. Therefore, noise and artifacts, which tend to increase the TV value, can be reduced via such regularization. For limited angle tomography, algorithms of adaptive steepest descent projection onto convex sets (ASD-POCS) \cite{sidky2006accurate,sidky2008image}, improved total variation (iTV) \cite{ritschl2011improved}, anisotropic total variation (aTV) \cite{chen2013limited}, reweighted total variation (wTV) \cite{huang2016image,huang2016watv}, and scale-space anisotropic total variation (ssaTV) \cite{huang2018scale} have been proposed. While TV methods achieve good reconstruction results when the missing angular range is small, they fail to reduce severe artifacts when a large angular range is missing. Moreover, they also require expensive computation and tend to lose high resolution details.

Recently, machine learning techniques have achieved overwhelming success in masses of fields including X-ray imaging. In the application of limited angle tomography, pixel-by-pixel artifact prediction using traditional machine learning is one direction \cite{huang2019traditional}. However, new artifacts might be introduced. Instead, deep learning methods have achieved impressive results. W\"urfl \textit{et al.} \cite{tobias2016deep,wurfl2018deep} proposed to learn certain weights based on known filtered back-projection (FBP) operators \cite{maier2019learning} to compensate missing data in limited angle tomography. Gu and Ye \cite{gu2017multi} proposed to learn artifacts from streaky images in a multi-scale wavelet domain using the U-Net architecture \cite{ronneberger2015u,falk2019u}. Bubba \textit{et al.} \cite{bubba2019learning} utilized an iterative shearlet transform algorithm to reconstruct visible singularities of an imaged object and a U-Net based neural network with dense blocks to predict invisible ones. In our previous work, we have demonstrated that deep learning is not robust to noise and adversarial examples \cite{huang2018some}. To improve image quality, a data consistent reconstruction method \cite{huang2019data} is proposed, where deep learning reconstruction is used as prior to provide information of missing data while conventional iterative reconstruction is applied to make deep learning reconstruction consistent to measured projection data.

In this work, deep learning is applied to limited angle reconstruction in the field of TXMs for the first time, to the best of our knowledge. Furthermore, training data is vital for deep learning methods. Without the access to real training data, in this work we will investigate the performance of deep learning trained from synthetic data.

\section{Materials And Method}

The proposed limited angle reconstruction method for TXMs consists of two steps: FBP preliminary reconstruction and deep learning reconstruction as post-processing. 

\subsection{FBP Preliminary Reconstruction}

For TXM systems with synchrotron radiation, parallel-beam X-rays are used. Each X-ray measures a line integral of the linear attenuation coefficients of a scanned sample, represented as,
\begin{equation}
\vp(u,v,\theta) = \iiint_{-\infty}^\infty \vf(x,y,z)\delta(x\cos\theta + y\sin \theta - u, z - v)\textrm{d}x\textrm{d}y\textrm{d}z,
\label{eqn:forwardProjection}
\end{equation}
where $\theta$ is the rotation angle of the sample, the rotation axis is parallel with the $z$-axis, $u$ and $v$ are the horizontal and vertical position indices at the detector respectively, $\vp(u,v,\theta)$ is the log-transformed projection, $\vf(x,y,z)$ is the attenuation distribution function of the sample, and $\delta(\cdot)$ is the Dirac delta function. 

In practice, noise always exists in measured projections due to various physical effects, e.\,g., Poisson noise. Since deep learning methods are sensitive to noise \cite{huang2018some}, noise reduction in input images is preferred. For this purpose, a penalized weighted least-square (PWLS) approach is utilized in projection domain. The objective function for PWLS is as follows \cite{wang2006penalized},
\begin{equation}
\Phi(\vp) = (\hat{\vp} - \vp)^\top \Lambda ^{-1}(\hat{\vp} - \vp) + \beta \cdot R(\vp),
\label{eqn:PWLS1}
\end{equation}
where $\vp$ is the vector of the ideal log-transformed projection, $\hat{\vp}$ is the vector of the measured log-transformed projection containing noise, $\vp_i$ is the $i^{\textbf{th}}$ element of $\vp$, $\Lambda$ is a diagonal matrix with the $i^{\textbf{th}}$ element equal to an estimate of the variance of $\hat{\vp}_i$, $R(\vp)$ is a regularization term, and $\beta$ is a relaxation parameter. The regularization term $R(\vp)$ is chosen as,
\begin{equation}
R(\vp) = \frac{1}{2}\sum_i\sum_{j\in\mathcal{N}_i} w_{i,j}(\vp_i - \vp_j)^2,
\label{eqn:PWLS2}
\end{equation}
where $\mathcal{N}_i$ is the 4-connectivity neighbourhood of the $i^{\textbf{th}}$ pixel and the weight $w_{i,j}$ is defined as,
\begin{equation}
w_{i,j} = \exp(-(\vp_i - \vp_j)^2/\sigma^2),
\label{eqn:PWLS3}
\end{equation}
with $\sigma$ a predefined parameter to control the weight.

The denoised projection is denoted by $\vp'(u,v,\theta)$. For image reconstruction, the filtered back-projection (FBP) algorithm with the Ram-Lak kernel $\vh(u)$ is applied,
\begin{equation}
\vf_{\text{FBP,PWLS}}(x,y,z)|_{z=v} = \int_{\theta_{\min}}^{\theta_{\max}}\int_{-\infty}^\infty\vp'(u,v,\theta) \vh(x\cos\theta + y\sin\theta-u) \textrm{d}u \textrm{d}\theta,
\label{eqn:backProjection}
\end{equation}
where $\theta_{\min}$ and $\theta_{\max}$ are the start rotation angle and the end rotation angle respectively and $\vf_{\text{FBP,PWLS}}$ is the FBP reconstruction from PWLS processed projection data. We further denote the FBP reconstruction from measured projection data without PWLS by $\vf_{\text{FBP}}$, i.e., replacing $\vp'(u,v,\theta)$ by $\hat{\vp}(u,v,\theta)$ in the above equation.

\subsection{Deep Learning Reconstruction}

\subsubsection{Neural network}

The above FBP reconstruction suffers from artifacts, typically in the form of streaks, due to missing data in limited angle tomography. To reduce artifacts, an image-to-image post-processing deep learning method using the U-Net is applied. 

\begin{figure}
\includegraphics[width = \linewidth]{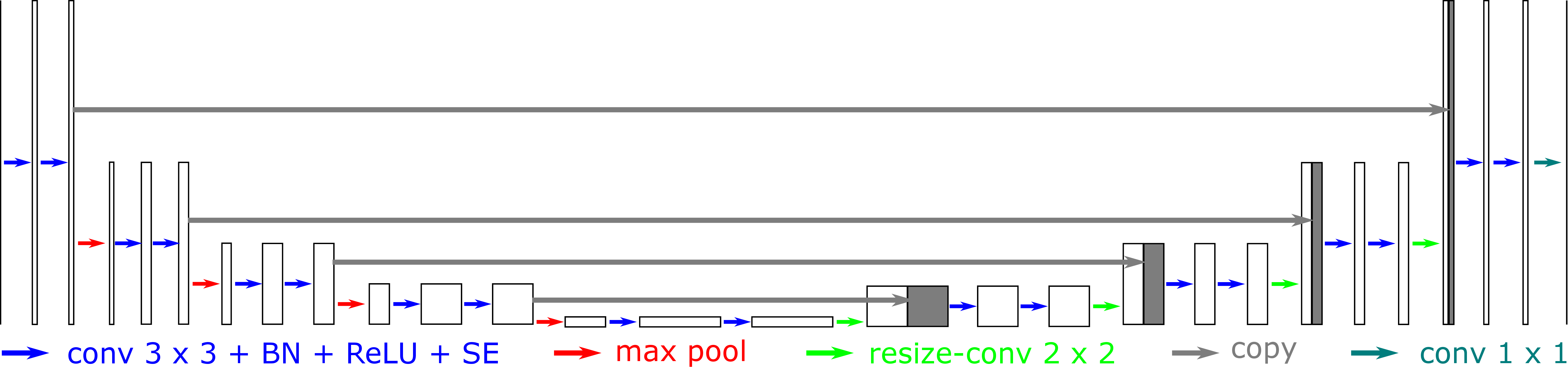}
\label{Fig:UNetArch}
\caption{The U-Net architecture for limited angle tomography.}
\end{figure}

The U-Net architecture for limited angle tomography is displayed in Fig.~\ref{Fig:UNetArch}. The input and output of the U-Net are both 2-D images of the same size. Each blue arrow stands for zero-padded $3 \times 3$ convolution followed by a rectified linear unit (ReLu), a batch normalization (BN) operation, and an squeeze-and-extraction (SE) block \cite{hu2018squeeze}. Each red arrow represents a max pooling operation to down-sample feature maps by a factor of 2. Each green arrow is a bilinear up-sampling operation followed by a $2 \times 2$ convolution to resize feature maps back. The grey arrows copy features from left side and concatenate them with the corresponding up-sampled features. The last $1\times 1$ convolution operation maps the multi-channel features to a desired output image. Because of the down/up-sampling and copy operations, the U-Net architecture has a large reception field and is able to learn features of multi-scales.

In this work, the input image is a 2-D horizontal slice from the FBP reconstruction without or with PWLS preprocessing, i.\,e., $\vf_{\text{FBP}}$ or $\vf_{\text{FBP,PWLS}}$ respectively. 
The output image is the corresponding artifact image. Hence, a final reconstruction of the U-Net, denoted by $\vfUNet$ or $\vfUNetP$ for the input image without and with PWLS respectively, is obtained by subtracting the input image by its corresponding predicted artifact image. For stable training, the input and output images are normalized to the range of [-1, 1] using the maximum intensity value of the input images.

Compared with the original U-Net architecture in \cite{ronneberger2015u}, the following modifications are made in the above U-Net architecture to improve its performance for limited angle tomography:

\begin{itemize}

\item \textbf{Zero-padded convolution:} In the original U-net architecture, unpadded convolution is used and the image size decreases after each convolution. Hence, information near image boundaries is missing in the output image. In this work, the zero-padded convolution is used to preserve image size. Because of this, the cropping operation is no longer necessary for each copy operation.

\item \textbf{Batch normalization:} The BN operation normalizes each convolutional layer's inputs in a mini-batch to a normal distribution with trained mean-shift and variance-scaling values. The BN technique allows neural networks to use higher learning rates and be less sensitive to initialization \cite{ioffe2015batch}. Therefore, it is a standard operation for convolutional neural networks nowadays.

\item \textbf{Squeeze-and-extraction:} The SE block \cite{hu2018squeeze} squeezes global spatial information into a channel descriptor by using global average pooling first. Afterwards, channel-wise dependencies are captured by a nonlinear excitation mechanism, which emphasizes multi-channel activations instead of single-channel activation. The SE technique adaptively recalibrates channel-wise feature responses to boost the representation power of a neural network.

\item \textbf{Resize and $2 \times 2$ convolution:} The original U-net architecture uses a deconvolution operation for up-sampling, which introduces checkerboard artifacts \cite{odena2016deconvolution}. To avoid this, we choose to resize each feature map using bilinear up-sampling with a scaling factor of 2 first. Afterwards, a $2 \times 2$ convolution operation is applied.

\item \textbf{Output and loss function:} 
The original U-Net is proposed for biomedical image segmentation, where the number of segmentation classes decides the channel number of the output image and each channel is a binary vector containing elements of 0 or 1. For segmentation, a softmax function is typically used to determine the highest probability class. Associated with the softmax activation in the output layer, the cross entropy loss function is typically used for training. As aforementioned, the output image is an 1-channel 2-D artifact image in this work. Therefore, the result of the $1 \times 1$ convolution is directly used as the output without any softmax function. Correspondingly, an $\ell_2$ loss function is used for training.
\end{itemize}

\subsubsection{Data preparation}
\label{subsubsect:dataPrepation}

 In order to reconstruct a sample from limited angle data using deep learning, training data is vital. However, on one hand it is very challenging to get a sufficient amount of real data; on the other hand, for most scans only limited angle data are acquired and hence reconstruction from complete data as ground truth is not available. Due to the scarcity of real data, we choose to train the neural network from synthetic data. For this purpose, two kinds of synthetic data are generated.

\textbf{Ellipsoid phantoms:} 3-D ellipsoid phantoms are designed, with two large ellipsoids to form an outer boundary, two middle-sized ellipsoids to simulate the cup-shaped chloroplast, 20 small ellipsoids to mimic lipid bodies, and 50 high intensity small-sized ellipsoids to simulate gold nanoparticles which are contained in the sample for geometry and motion calibration \cite{wang2019jitter}. The locations, sizes, and intensities of the ellipsoids are randomly generated. Since many samples are immobilized in a certain condition, e.\,g. in an ice tube in this work, a background with a constant intensity of 0.002\,\textmu m$^{-1}$ is added.

\textbf{Multi-category data:} For a certain parallel-beam limited angle tomography system, no matter what kinds of objects are imaged, the projections and the FBP reconstructions should follow the mathematics in Eqns.~(\ref{eqn:forwardProjection}) and (\ref{eqn:backProjection}). In addition, based on the theories of transfer learning \cite{pan2009survey}, one/zero-shot learning \cite{fei2006one,palatucci2009zero}, a neural network trained for one task can also generalize to another similar task. Therefore, in this work, images of multi-categories are collected to train the neural network for complex structures, for example, optical microscopy algae images and medical CT images. Note that although TXMs data for chlorella cells, the test sample in this work, are not accessible, data of algae cells in other imaging modalities, especially in optical microscopies, are abundant. Images in other modalities also share a plenty of useful structure information as TXMs do.

\subsection{Experimental Setup}

\subsubsection{Synthetic Data}

For deep learning training, 10 ellipsoid phantoms with a size of $512\times 512 \times 512$ are generated. From each 3-D phantom, 20 slices are uniformly selected. From the multi-category data, 400 image slices are collected. Color images are converted to grey intensity images. The above images are further rotated by $90^\circ$, $180^\circ$, and $270^\circ$. Therefore, 2400 image slices in total are synthesized for training. 

\begin{figure}
\centering
\includegraphics[width = 150pt]{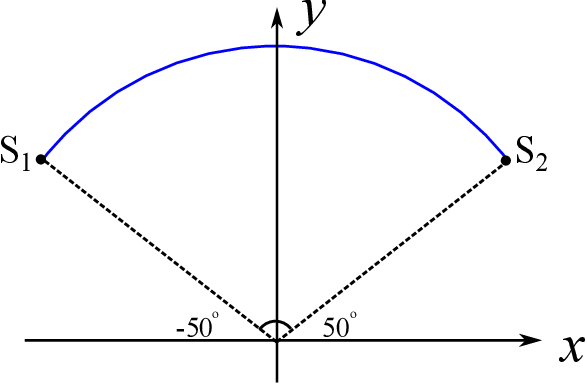}
\caption{The scanned angular range of the TXM system is from $-50^\circ$ to $50^\circ$.}
\label{Fig:geometry}
\end{figure}

Parallel-beam sinograms are simulated from rotation angle $-50^\circ$ to $50^\circ$ with an angular step of $1^\circ$, as displayed in Fig.~\ref{Fig:geometry}. The detector size is 512 with a pixel size of 21.9\,nm. To improve the robustness of the neural network to noise, Poisson noise is simulated considering a photon number of $10^4$, $5.0\times10^4$, or $10^5$ for each X-ray before attenuation. Here multiple dose levels are used to improve the robustness of the neural network to different levels of noise. For training, 1200 preliminary image slices with a size of $256 \times 256$ is reconstructed by FBP using the Ram-Lak kernel directly from noisy projection data for the 600 original slices and their $90^\circ$ rotations, while the other 1200 slices are reconstructed from projection data processed by 2 iterations of PWLS. To obtain the diagonal matrix $\Lambda$ in Eqn.~(\ref{eqn:PWLS1}), the variance of each detector pixel $\hat{\vp}_i$ is estimated by the following formula \cite{wang2006penalized}, 
\begin{equation}
\sigma^2_i = a_i\cdot\exp{(\hat{\vp}_i/\eta)},
\end{equation}
where $a_i$ is set to 0.5 for each pixel $i$ and $\eta$ is set to 1. The value of $\sigma$ in Eqn.~(\ref{eqn:PWLS3}) is set to 2.

The U-Net is trained on the above synthetic data using the Adam optimizer for 500 epochs. The learning rate is $10^{-3}$ for the first 100 epochs and gradually decreases to $10^{-5}$ for the last epochs. The $\ell_2$-regularization with a parameter of $10^{-4}$ is applied to avoid large network weights.

For a preliminary quantitative evaluation, the trained U-Net model is evaluated on one new synthetic ellipsoid phantom first. Its limited angle projection data are generated with Poisson noise using a photon number of $10^4$. The projections are denoised by 2 iterations of PWLS. 

\subsubsection{Chlorella Data}

\begin{figure}
\centering
\includegraphics[width = 0.5\linewidth]{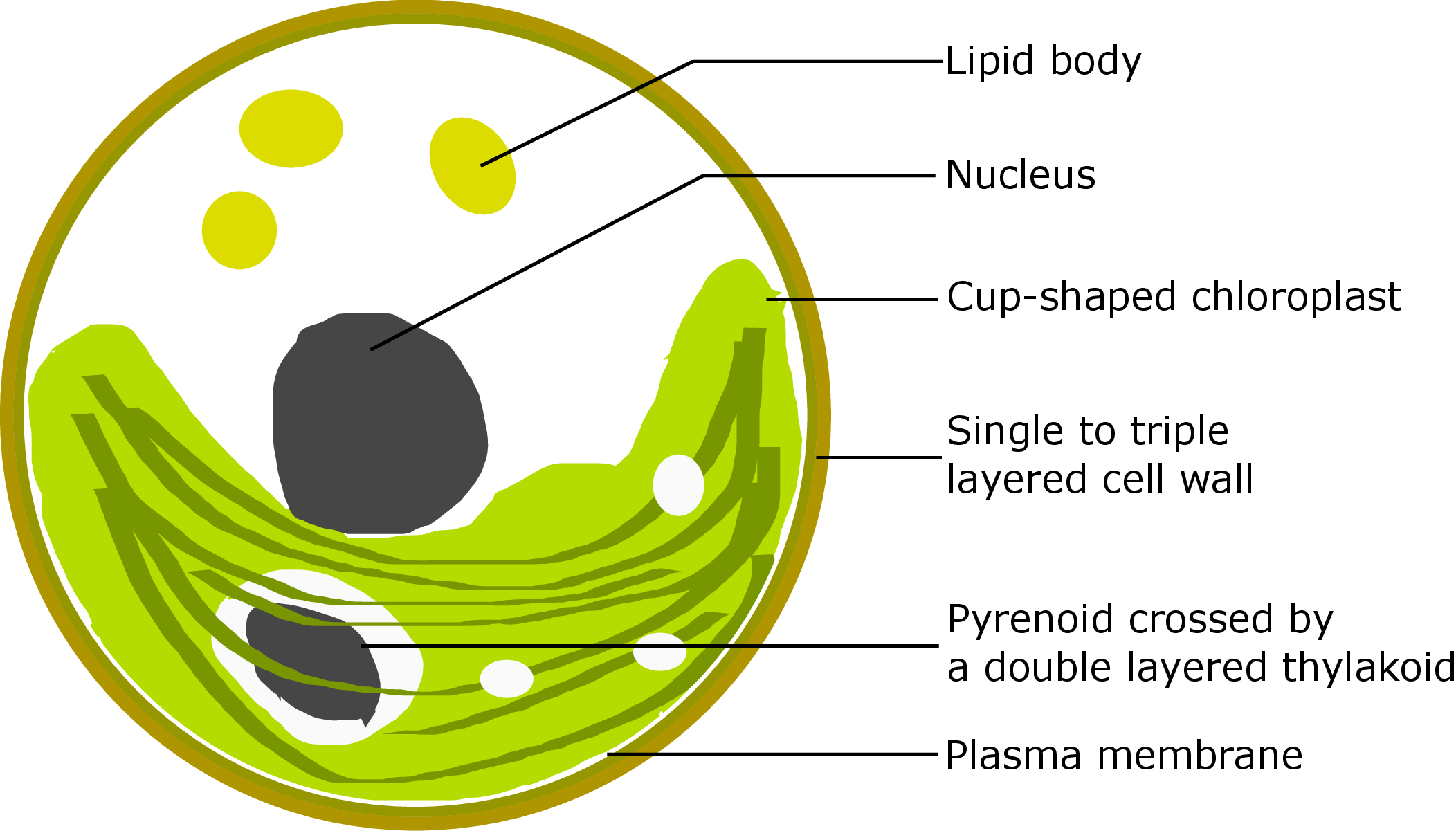}
\caption{The main structures of a chlorella cell \cite{baudelet2017new}.}
\label{Fig:Chlorella}
\end{figure}

As a demonstration example, a sample of chlorella cells is scanned in a soft X-ray microscope at beamline BL07W \cite{liu2018quantitative} in the National Synchrotron Radiation Laboratory (NSRL) in Hefei, China. Chlorella is a genus of single-celled green algae with a size of 2\,\textmu m to 10\,\textmu m. It mainly consists of a single to triple layered cell wall, a thin plasma membrane, a nucleus, a cup-shaped chloroplast, a pyrenoid, and several lipid bodies, as illustrated in Fig.~\ref{Fig:Chlorella} \cite{baudelet2017new}.

To hold the chlorella sample, a traditional 100-mesh transmission electron microscopy (TEM) grid is used. Because of the TEM grid, a valid scan of $100^\circ$ ($-50^\circ$ to $50^\circ$ in Fig.~\ref{Fig:geometry} with an angular step of $1^\circ$) only is acquired to avoid collision between the grid and the zone plate. Rapid freezing of the chlorella sample with liquid nitrogen is performed before scanning to immobilize the cells in an ice tube and suppress the damage of radiation to cellular structures. The X-ray energy used in the experiment is 520\,eV for the so-called ``water window". Each projection image is rebinned to a size of $512 \times 512$ with a pixel size of 21.9\,nm $\times$ 21.9\,nm. As the shift of rotation axis \cite{yang2015registration} and jitter motion \cite{yu2018automatic} are two main causes of image blurry, they are solved via measurement of geometric moments after acquisition, as described in \cite{wang2019jitter}. The projections are denoised by 2 iterations of PWLS afterwards. 

\section{Results And Discussion}

\subsection{Ellipsoid Phantom Results}

\begin{figure}

\centering

\begin{minipage}{0.23\linewidth}
{
\includegraphics[width=\linewidth]{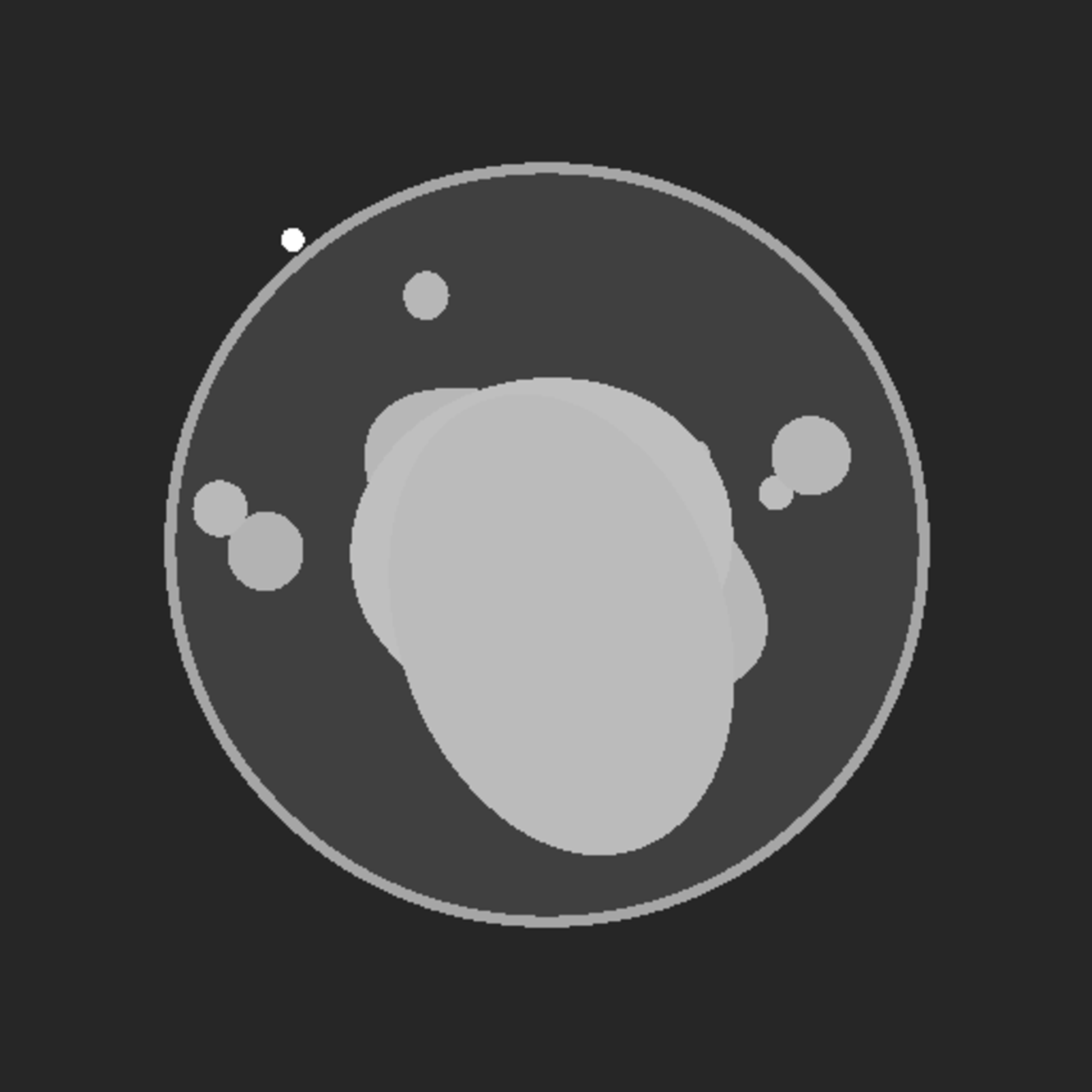}
}
\centerline{(a) $\vf_{\text{Referece}}$}
\end{minipage}
\hspace{5pt}
\begin{minipage}{0.23\linewidth}
{
\includegraphics[width=\linewidth]{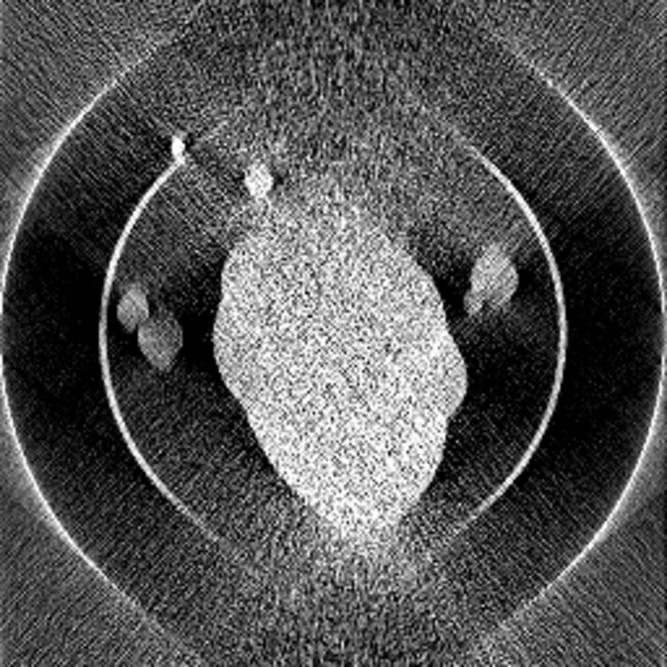}
}
\centerline{(b) $\vf_{\text{FBP}}$, 3.61}
\end{minipage}
\hspace{5pt}
\begin{minipage}{0.23\linewidth}
{
\includegraphics[width=\linewidth]{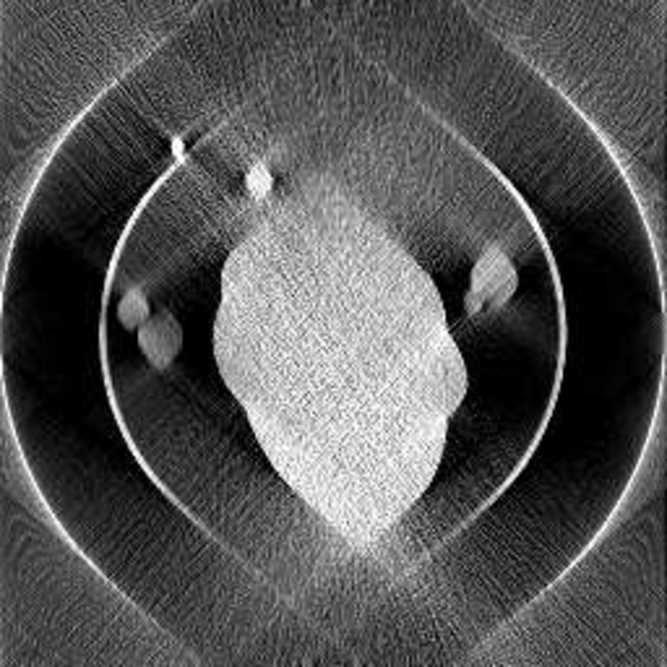}
}
\centerline{(c) $\vf_{\text{FBP, PWLS}}$, 3.45}
\end{minipage}

\begin{minipage}{0.23\linewidth}
{
\
}
\end{minipage}
\hspace{5pt}
\begin{minipage}{0.23\linewidth}
{
\includegraphics[width=\linewidth]{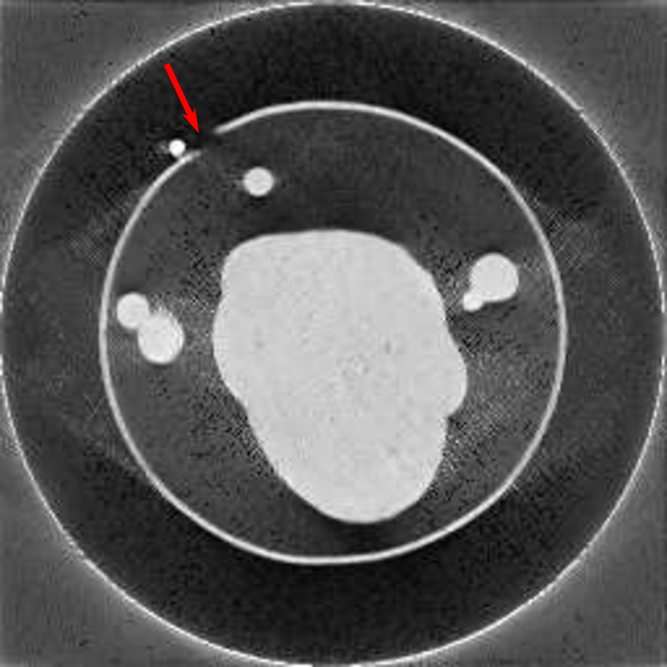}
}
\centerline{(d) $\vfUNet$, 1.65}
\end{minipage}
\hspace{5pt}
\begin{minipage}{0.23\linewidth}
{
\includegraphics[width=\linewidth]{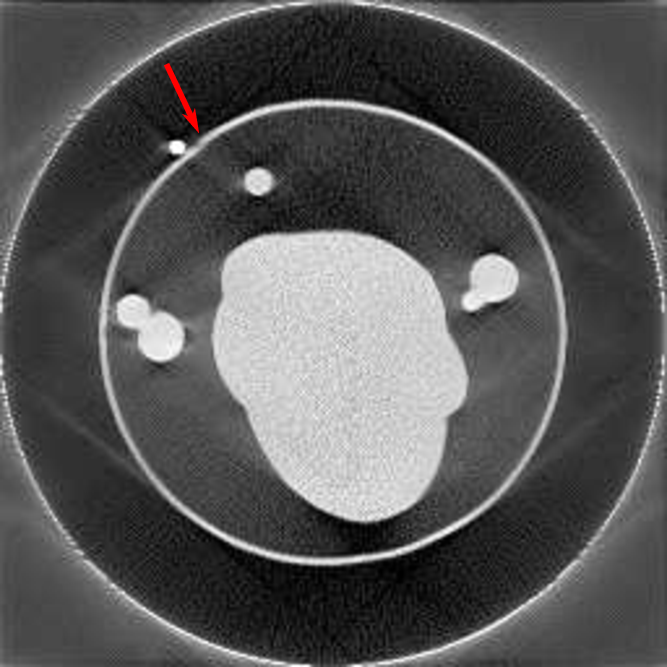}
}
\centerline{(e) $\vfUNetP$, 1.58}
\end{minipage}

\caption{The reconstruction results of the $215^{\text{th}}$ slice in the test ellipsoid phantom without and with PWLS using a photon number of $10^4$, window [0, 0.02]\,\textmu m$^{-1}$. The RMSE of each image w.\,r.\,t.~the corresponding reference is displayed at the subcaption with the unit $10^{-3}$\,\textmu m$^{-1}$. The boundary indicated by the red arrow is disconnected in $\vfUNet$, while it is reconstructed in $\vfUNetP$.}
\label{Fig:phantomNoiseResults}
\end{figure}

The reconstruction results without and with PWLS for the $250^{\text{th}}$ slice of the test ellipsoid phantom using a photon number of $10^4$ are displayed in Fig.~\ref{Fig:phantomNoiseResults}. The root-mean-square error (RMSE) inside the field-of-view (FOV) of each image slice with respect to (w.\,r.\,t.) the corresponding reference slice is displayed in the subcaption. In Figs.~\ref{Fig:phantomNoiseResults}(b)-(e), the outer ring is caused by the lateral truncation and it is preserved to mark the FOV.

 The FBP reconstruction from $100^\circ$ limited angle data without PWLS preprocessing, $\vf_{\text{FBP}}$, is displayed in Fig.~\ref{Fig:phantomNoiseResults}(b). Compared with the reference image $\vf_\text{Reference}$, only the structures with an orientation inside the scanned angular range (Fig.~\ref{Fig:geometry}) are reconstructed while all other structures are severely distorted. In addition, the Poisson noise pattern is clearly observed due to the low dose. In contrast, Poisson noise is prominently reduced by PWLS in $\vf_{\text{FBP,PWLS}}$, as displayed in Fig.~\ref{Fig:phantomNoiseResults}(c). The U-Net reconstruction with the input of $\vf_{\text{FBP}}$ is displayed in Fig.~\ref{Fig:phantomNoiseResults}(d), where most ellipsoid boundaries are restored well. The RMSE inside the FOV is reduced from $3.61 \times 10^{-3}$\,\textmu m$^{-1}$ in $\vf_{\text{FBP}}$ to $1.65 \times 10^{-3}$\,\textmu m$^{-1}$ in $\vfUNet$. This demonstrates the efficacy of deep learning in artifact reduction for limited angle tomography. However, some Poisson noise remains in Fig.~\ref{Fig:phantomNoiseResults}(d). Especially, the boundary indicated by the red arrow is disconnected in $\vfUNet$. The U-Net reconstruction with the input of $\vf_{\text{FBP,PWLS}}$ is displayed in Fig.~\ref{Fig:phantomNoiseResults}(e), achieving the smallest RMSE value of $1.58 \times 10^{-3}$\,\textmu m$^{-1}$. Importantly, the disconnected boundary fragment indicated by the red arrow is reconstructed in $\vfUNetP$. This demonstrates the benefit of PWLS preprocessing.

The average RMSE and structural similarity (SSIM) index of all slices in the FBP and U-Net reconstructions without and with PWLS for the test ellipsoid phantom are displayed in Table~\ref{tab:RMSE}. The U-Net reduces the average RMSE value from $2.55 \times 10^{-3}$\,\textmu m$^{-1}$ in $\vf_{\text{FBP}}$ to $1.21 \times 10^{-3}$\,\textmu m$^{-1}$ in $\vfUNet$. With PWLS, the average RMSE is further reduced to $1.16 \times 10^{-3}$\,\textmu m$^{-1}$ in $\vfUNetP$. Consistently, $\vfUNetP$ achieves a larger SSIM index than $\vfUNet$. This quantitative evaluation also demonstrates the efficacy of the U-Net in limited angle tomography and the benefit of PWLS preprocessing.

\begin{table}
\caption{The average RMSE and SSIM values for each reconstruction method using a photon number of $10^4$ without or with PWLS, unit for RMSE: $10^{-3}$\textmu m$^{-1}$.}
\label{tab:RMSE}
\begin{tabular}{lcccc}
Metric   &$\vf_{\text{FBP}}$    &$\vf_{\text{FBP, PWLS}}$    &$\vfUNet$   &$\vfUNetP$   \\
\hline
RMSE &2.55 &2.44 &1.21 & \textbf{1.16} \\
\hline
SSIM &0.625 &0.648 &0.920 & \textbf{0.932} \\
\hline
\end{tabular}
\end{table}

\subsection{Chlorella Results}

\begin{figure}
\centering
\begin{minipage}{0.3\linewidth}
{
\includegraphics[width=\linewidth]{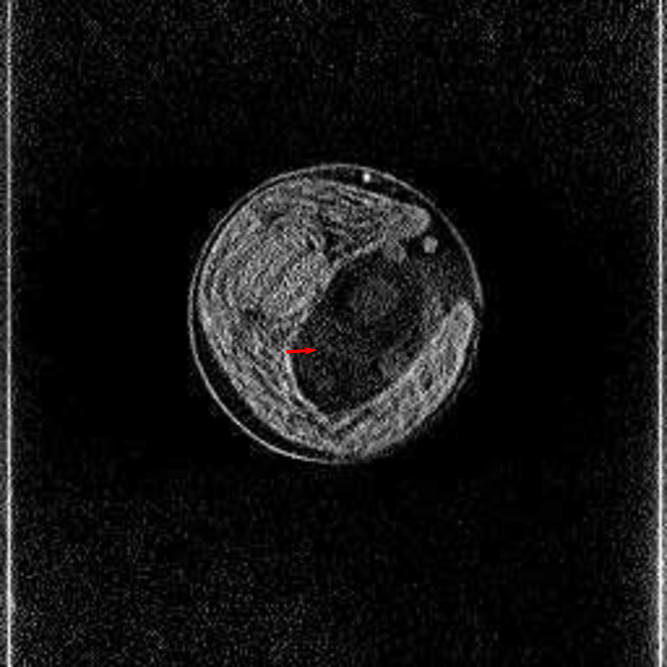}
}
\centerline{(a) $\vf_{\text{sag, FBP}}$}
\end{minipage}
\hspace{5pt}
\begin{minipage}{0.3\linewidth}
{
\includegraphics[width=\linewidth]{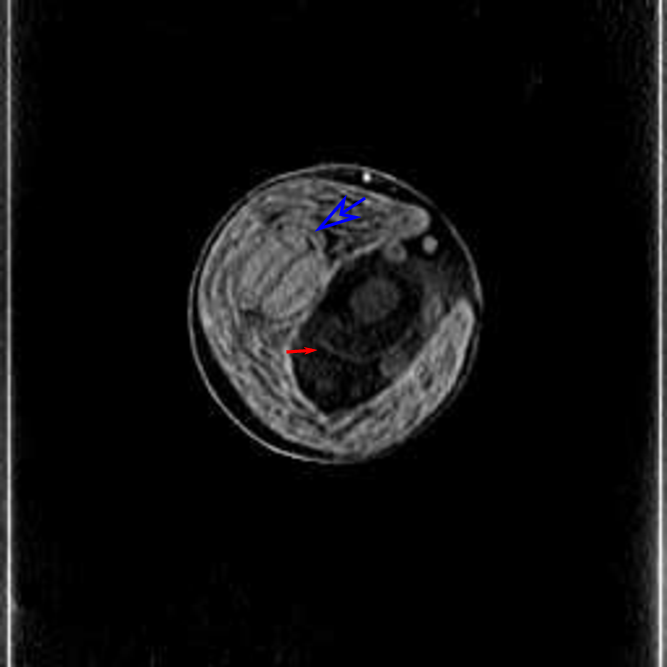}
}
\centerline{(b) $\vf_{\text{sag, FBP, PWLS}}$}
\end{minipage}
\caption{The $103^{\text{th}}$ slice from the sagittal view reconstructed from projections without and with PWLS preprocessing. The nucleus membrane in (a) and (b) is indicated by the red solid arrow. The pyrenoid membrane is indicated by the blue hollow arrow. Window: [0, 0.015]\,\textmu m$^{-1}$. }
\label{Fig:tomosynsthesisSlices}
\end{figure}

To demonstrate the benefit of PWLS for the chlorella data, horizontal slices are reconstructed by FBP from the chlorella projection data without or with PWLS processing. A 3-D volume is obtained by stacking the horizontal slices. Sagittal slices are obtained by reslicing the volume into 256 slices in the sagittal view. The sagittal slices from projections without and with PWLS are denoted by $\vf_{\text{sag, FBP}}$ and $\vf_{\text{sag, FBP, PWLS}}$, respectively. The results of the $103^{\text{th}}$ slice are displayed in Fig.~\ref{Fig:tomosynsthesisSlices}. Fig.~\ref{Fig:tomosynsthesisSlices}(a) exhibits that the subcellular structures of cell wall, chloroplast, lipid bodies, nucleus, and pyrenoid are reconstructed. However, due to noise, the nucleus membrane is barely seen, which is indicated by the red solid arrow. In contrast, with PWLS, the nucleus membrane is observed better, as indicated by the red solid arrow in Fig.~\ref{Fig:tomosynsthesisSlices}(b). Moreover, the textures in the cup-shaped chloroplast are also observed better in Fig.~\ref{Fig:tomosynsthesisSlices}(b) than those in Fig.~\ref{Fig:tomosynsthesisSlices}(a). For example, the pyrenoid membrane inside the chloroplast is well observed, as indicated by the blue hollow arrow in Fig.~\ref{Fig:tomosynsthesisSlices}(b). These observations demonstrate the benefit of PWLS. 

\begin{figure}
\centering

\begin{minipage}{0.23\linewidth}
{
\includegraphics[width=\linewidth]{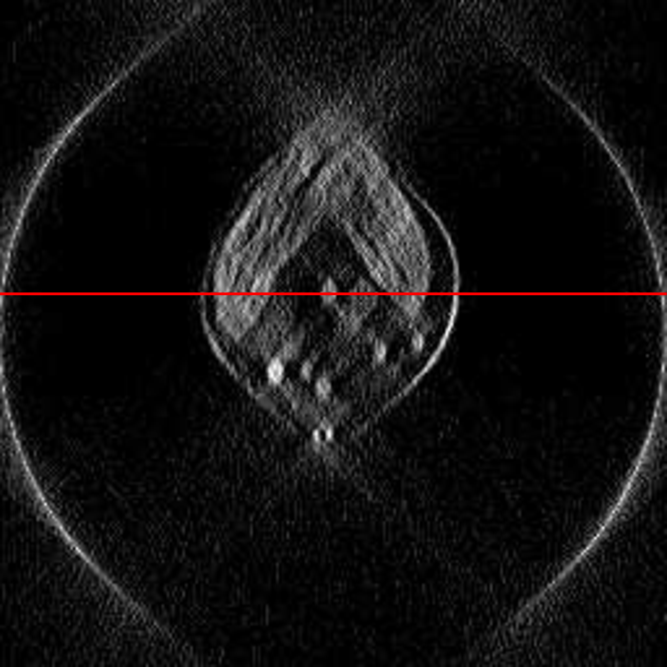}
}
\centerline{(a) $\vf_{\text{FBP}}$}
\end{minipage}
\begin{minipage}{0.23\linewidth}
{
\includegraphics[width=\linewidth]{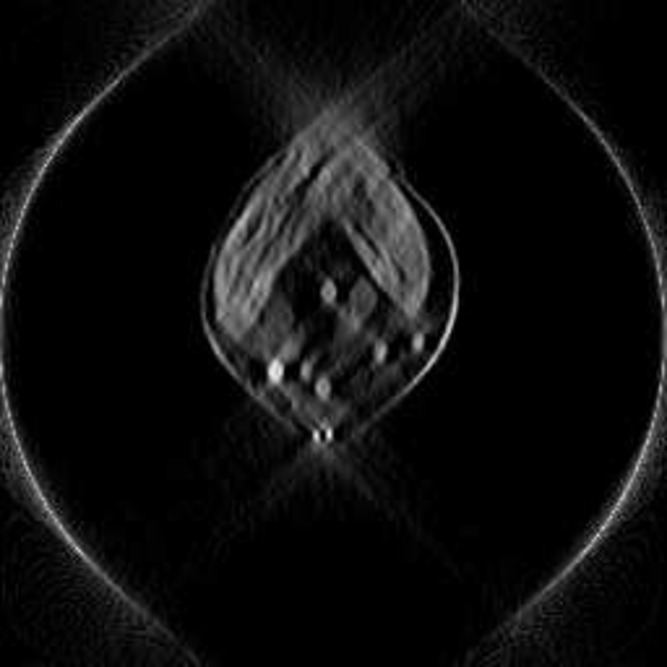}
}
\centerline{(b) $\vf_{\text{FBP, PWLS}}$}
\end{minipage}
\begin{minipage}{0.23\linewidth}
{
\includegraphics[width=\linewidth]{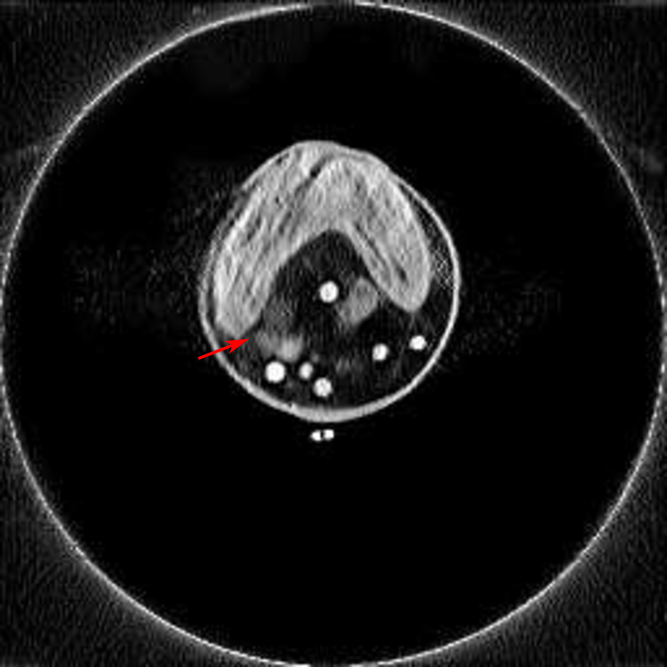}
}
\centerline{(c) $\vfUNet$}
\end{minipage}
\begin{minipage}{0.23\linewidth}
{
\includegraphics[width=\linewidth]{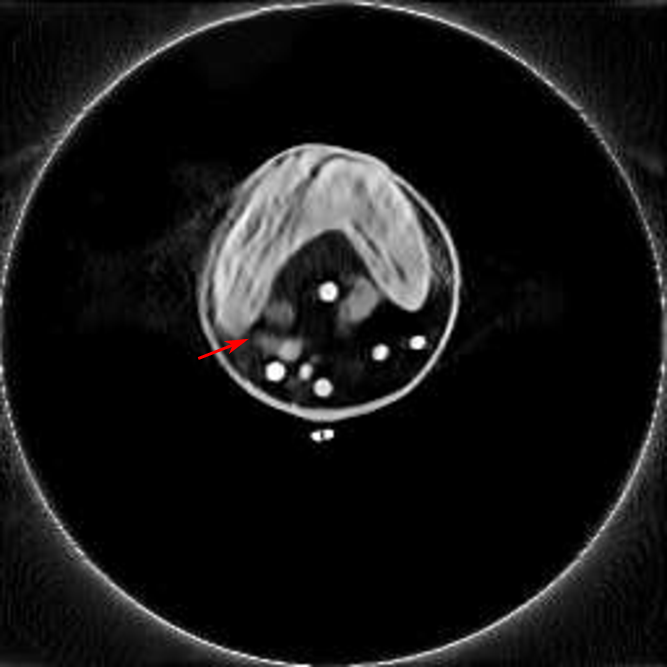}
}
\centerline{(d) $\vfUNetP$}
\end{minipage}

\begin{minipage}{0.23\linewidth}
{
\includegraphics[width=\linewidth]{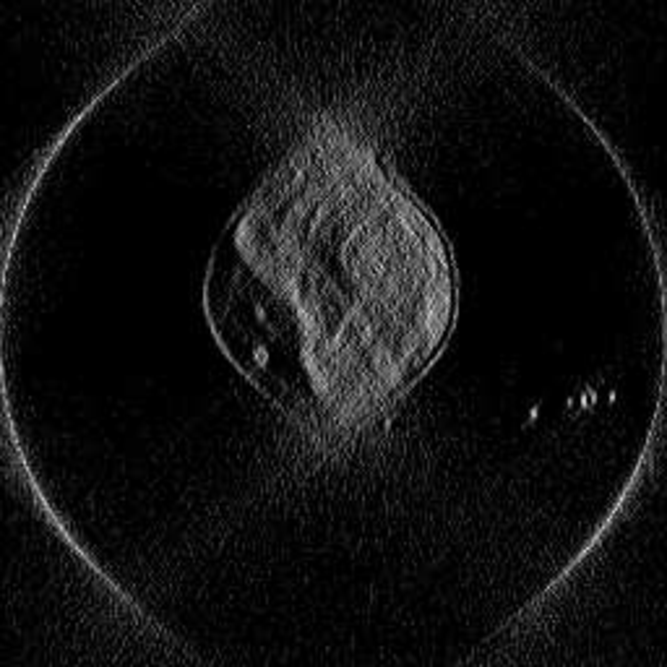}
}
\centerline{(e) $\vf_{\text{FBP}}$}
\end{minipage}
\begin{minipage}{0.23\linewidth}
{
\includegraphics[width=\linewidth]{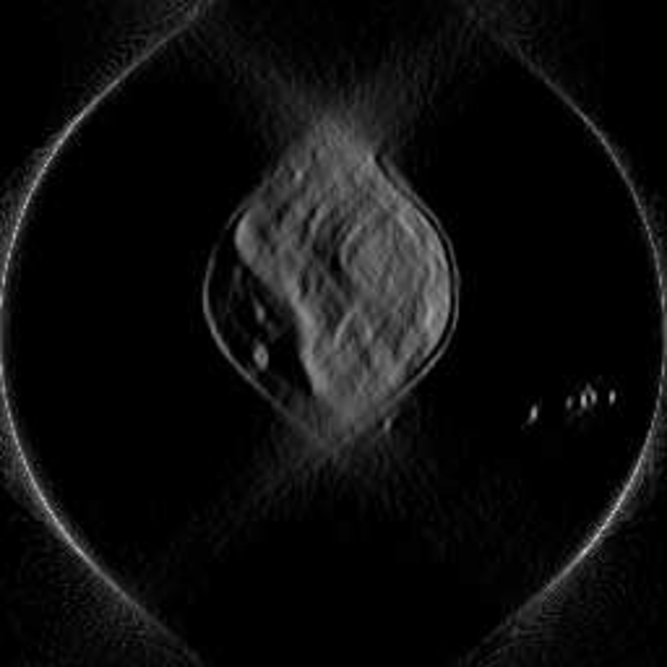}
}
\centerline{(f) $\vf_{\text{FBP, PWLS}}$}
\end{minipage}
\begin{minipage}{0.23\linewidth}
{
\includegraphics[width=\linewidth]{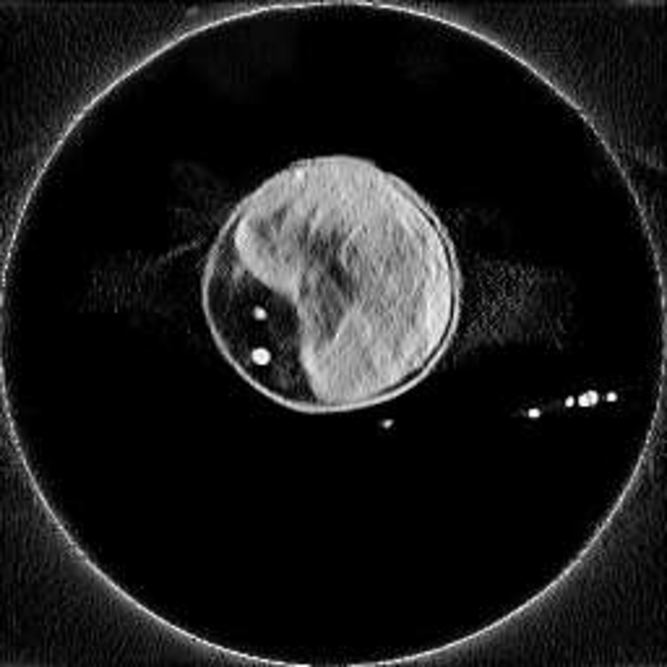}
}
\centerline{(g) $\vfUNet$}
\end{minipage}
\begin{minipage}{0.23\linewidth}
{
\includegraphics[width=\linewidth]{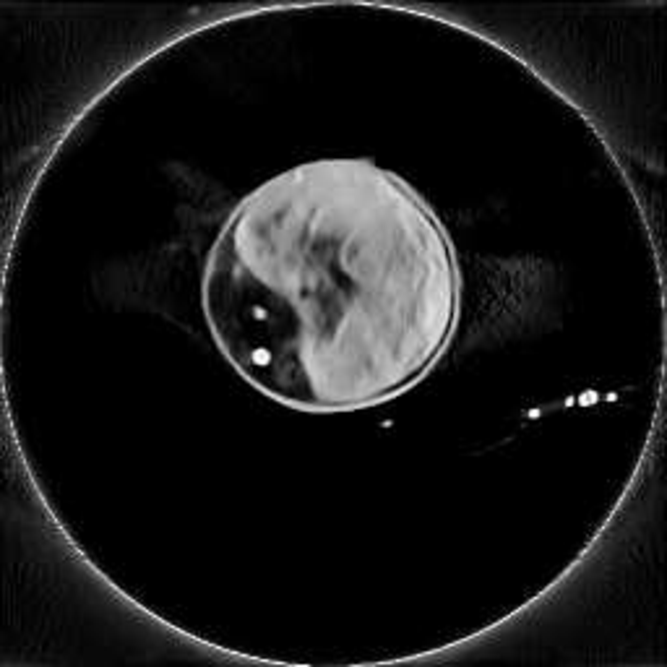}
}
\centerline{(h) $\vfUNetP$}
\end{minipage}
\caption{The reconstruction results of two horizontal slices for the chlorella data, window: [0.003, 0.015]\,\textmu m$^{-1}$. The top row is for the $213^{\text{th}}$ slice while the bottom row is for the $331^{\text{th}}$ slice. The red line in (a) indicates the position of intensity profiles in Fig.~\ref{Fig:lineProfiles}.}
\label{Fig:chlorellaResults}
\end{figure}

The reconstruction results of two horizontal example slices are displayed in Fig.~\ref{Fig:chlorellaResults}. 
Figs.~\ref{Fig:chlorellaResults}(a) and (b) are FBP reconstruction images of the $213^{\text{th}}$ slice without and with PWLS respectively, where many subcellular structures of the chlorella, e.\,g. the cell wall, chloroplast and lipid bodies, are severely distorted. Compared with Fig.~\ref{Fig:chlorellaResults}(a), Fig.~\ref{Fig:chlorellaResults}(b) contains less noise due to PWLS preprocessing. Their corresponding deep learning results $\vfUNet$ and $\vfUNetP$ are displayed in Figs.~\ref{Fig:chlorellaResults}(c) and (d), respectively. The cell walls are restored and the chloroplasts exhibit a good ``C" shape in both images. In addition, the lipid bodies and the gold nanoparticles are well observed. These observations demonstrate the efficacy of deep learning for limited angle tomography on real data. Moreover, the lipid bodies indicated by the arrows in Fig.~\ref{Fig:chlorellaResults}(d) are separated better than those in Fig.~\ref{Fig:chlorellaResults}(c), which highlights the benefit of PWLS preprocessing for deep learning reconstruction. 

For the reconstruction results of the $331^{\text{th}}$ slice displayed in the bottom row, the U-Net is also able to reconstruct the cell wall, the chloroplast, and lipid bodies. With PWLS, $\vfUNetP$ in Fig.~\ref{Fig:chlorellaResults}(h) contains less noise than $\vfUNet$ in Fig.~\ref{Fig:chlorellaResults}(g), consistently demonstrating the benefit of PWLS.

\begin{figure}

\centering

\begin{minipage}{0.49\linewidth}
{
\includegraphics[width=\linewidth]{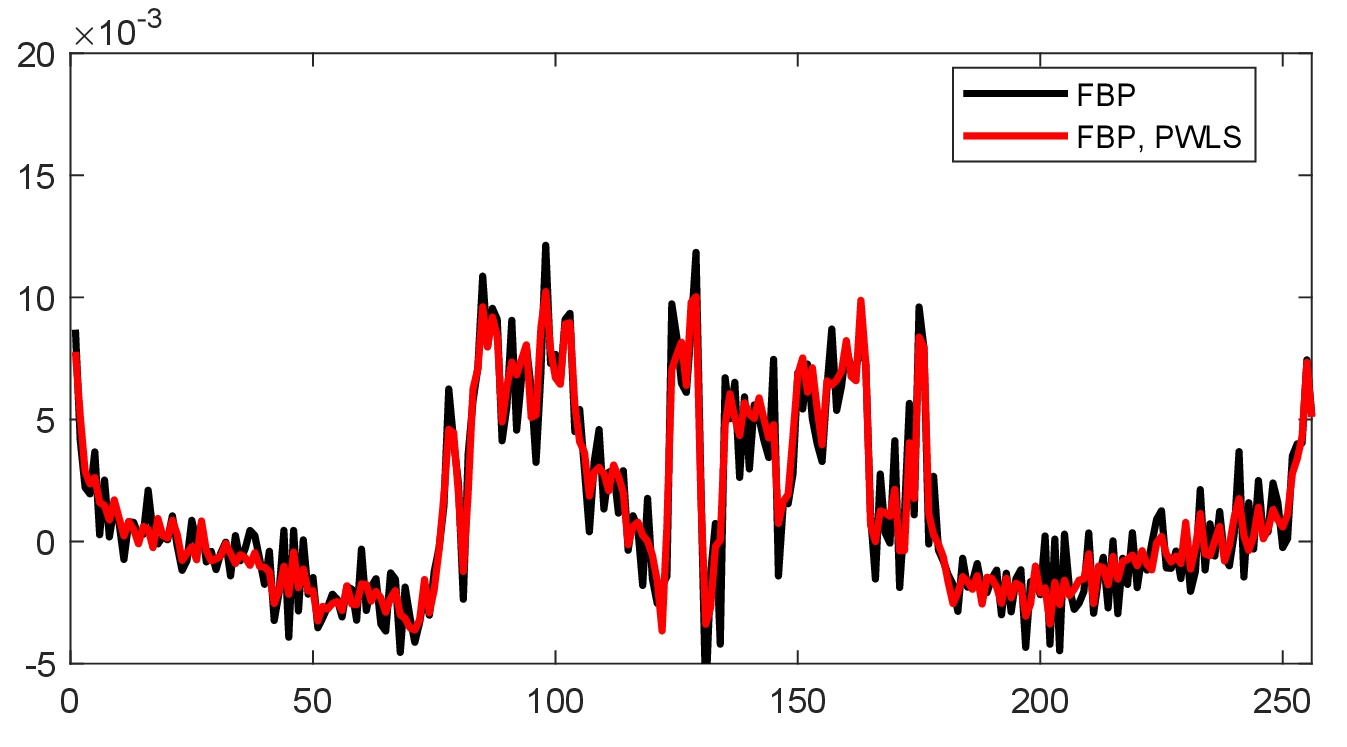}
}
\centerline{(a) Line profiles of FBP reconstructions}
\end{minipage}
\begin{minipage}{0.49\linewidth}
{
\includegraphics[width=\linewidth]{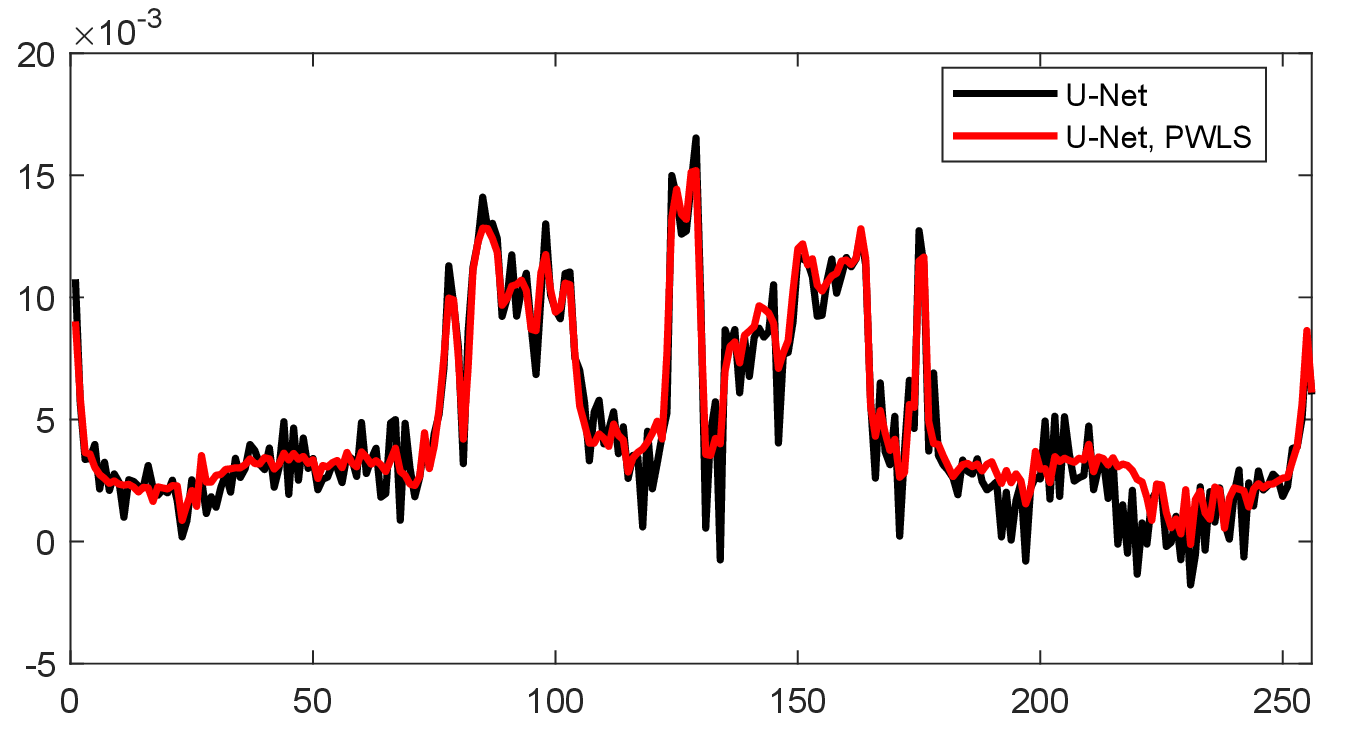}
}
\centerline{(b) Line profiles of U-Net reconstructions}
\end{minipage}

\caption{The intensity profiles of a line in the FBP and U-Net reconstructions without and with PWLS. The position of the line is indicated in Fig.~\ref{Fig:chlorellaResults}(a).}
\label{Fig:lineProfiles}
\end{figure}

For image quality quantification, the intensity profiles of a line in the FBP and U-Net reconstructions without and with PWLS are displayed in Fig.~\ref{Fig:lineProfiles}. The position of the line is indicated in Fig.~\ref{Fig:chlorellaResults}(a). In Fig.~\ref{Fig:lineProfiles}(a), the line profiles of $\vf_{\text{FBP}}$ and $\vf_{\text{FBP, PWLS}}$ are displayed. For both profiles, in the pixels of 0-70 and 180-256, the intensity value increases from the center outward, which is a characteristic of cupping artifacts and indicates the existence of data truncation. In the profile of $\vf_{\text{FBP}}$, a lot of high frequency oscillations are observed, while many of them are mitigated in $\vf_{\text{FBP, PWLS}}$ by PWLS. In Fig.~\ref{Fig:lineProfiles}(b), high frequency oscillations are observed in the profile of $\vfUNet$ as well, while the profile of $\vfUNetP$ has relatively smooth transitions. This demonstrates the benefit of PWLS in avoiding high frequency noise in the U-Net reconstruction.

\begin{figure}
\centering

\begin{minipage}{0.3\linewidth}
{
\includegraphics[width=\linewidth]{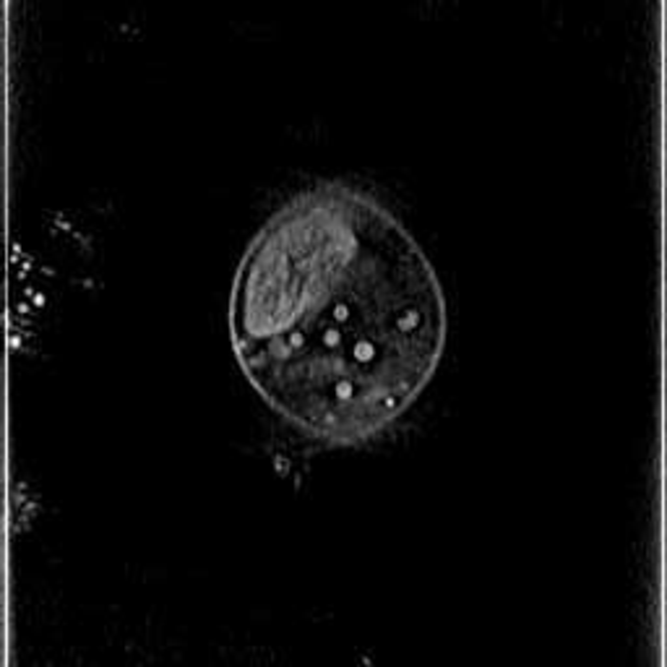}
}
\centerline{(a) $\vf_{\text{sag, PWLS}}$}
\end{minipage}
\hspace{5pt}
\begin{minipage}{0.3\linewidth}
{
\includegraphics[width=\linewidth]{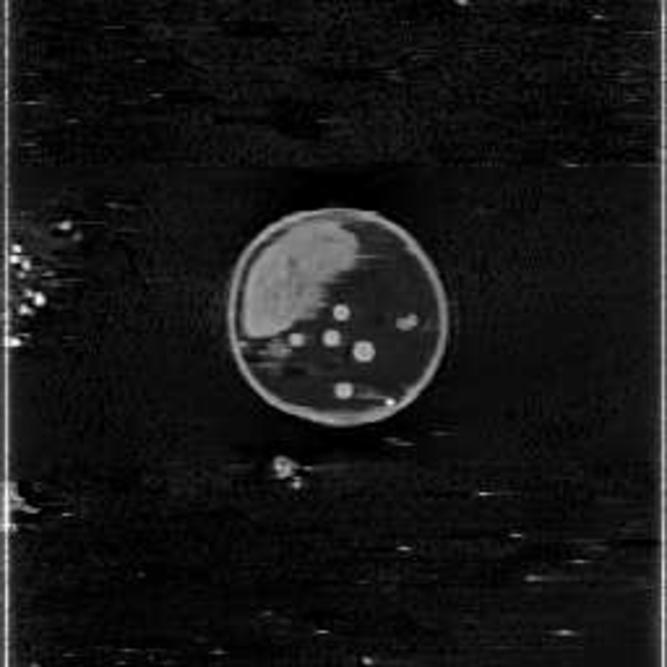}
}
\centerline{(b) $\vf_{\text{sag, U-Net, PWLS}}$}
\end{minipage}

\caption{The $150^{\text{th}}$ slices from the sagittal view reconstructed by FBP and U-Net with PWLS preprocessing, window: [0, 0.015]\,\textmu m$^{-1}$. }
\label{Fig:tomosynsthesisSlice150}
\end{figure}

In the sagittal view, although structures are observed well for central slices such as the $103^{\text{th}}$ slice, structures in many other slices are distorted due to missing data. For example, the $150^{\text{th}}$ sagittal slice of the FBP reconstruction $\vf_\text{FBP, PWLS}$ is displayed in Fig.~\ref{Fig:tomosynsthesisSlice150}(a), where the cell wall is severely distorted. With the proposed U-Net reconstruction with PWLS preprocessing, the cell wall is restored in an approximate round shape, as shown in Fig.~\ref{Fig:tomosynsthesisSlice150}(b).

\begin{figure}
\centering

\begin{minipage}{0.45\linewidth}
{
\includegraphics[width=\linewidth]{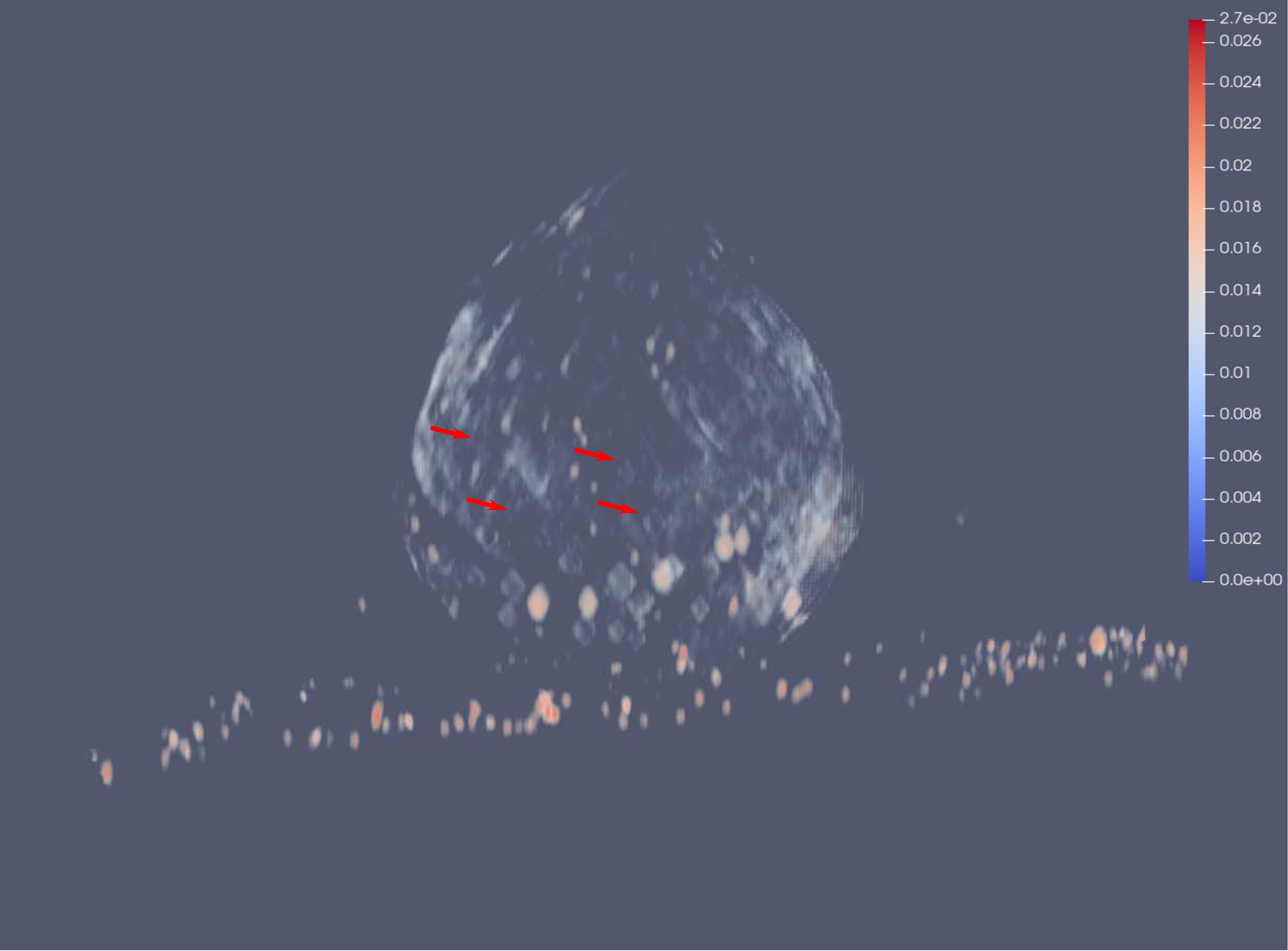}
}
\centerline{(a) $\vf_{\text{FBP, PWLS}}$}
\end{minipage}
\begin{minipage}{0.45\linewidth}
{
\includegraphics[width=\linewidth]{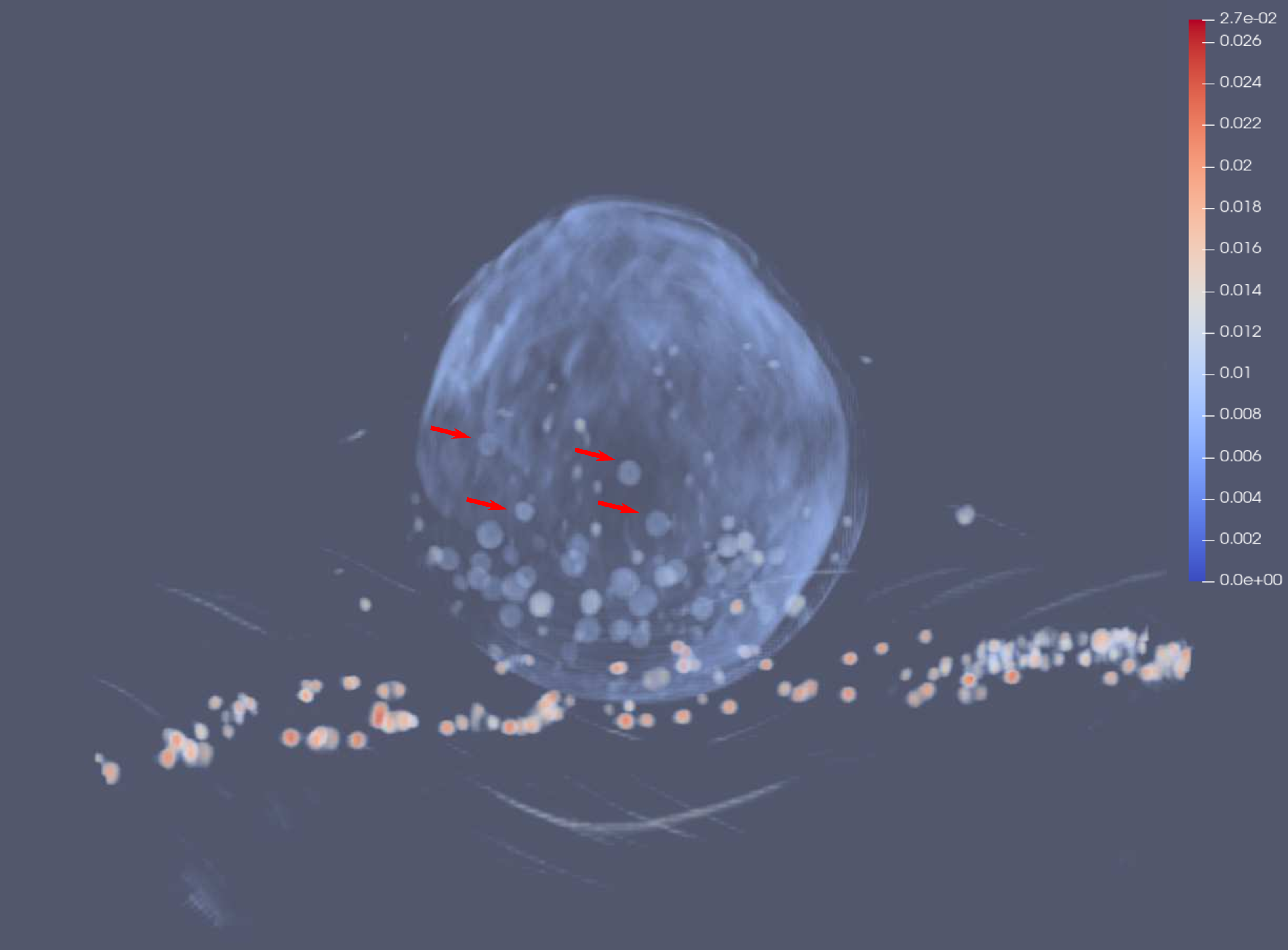}
}
\centerline{(b) $\vfUNetP$}
\end{minipage}
\caption{The 3-D rendering of the volumes reconstructed by FBP and U-Net respectively with PWLS using the tool of ParaView, which is viewed along the $z$ direction. The lipids indicated by the arrows are observed well in $\vfUNetP$ while they are barely seen in $\vf_{\text{FBP, PWLS}}$.}
\label{Fig:paraView}
\end{figure}

The volumes reconstructed by FBP and U-Net with PWLS are rendered by ParaView, an open-source 3-D visualization tool, and displayed in Figs.~\ref{Fig:paraView}(a) and (b), respectively. Fig.~\ref{Fig:paraView}(a) displays that the top and bottom parts of the chlorella cell are missing. In addition, the shapes of lipid bodies are distorted. Instead, the top and bottom parts are restored by the U-Net. Regarding the lipid bodies, their shapes are also restored to round shapes. Moreover, in the U-Net reconstruction, the lipid bodies indicated by the arrows are observed well while they are barely seen in the FBP reconstruction. This 3-D rendering result highlights the benefit of U-Net in the 3-D visualization of subcellular structures.

\subsection{Discussion}

As a state-of-the-art method, the U-Net achieves significant improvement in image quality from the FBP reconstructions, achieving the best average RMSE value in Table~\ref{tab:RMSE}. However, in some cases, the structures it predicts are not accurate. For example, the cell wall is not in a perfect round shape in Fig.~\ref{Fig:chlorellaResults}(d) and Fig.~\ref{Fig:tomosynsthesisSlice150}(b). This is potentially caused by various factors such as noise, insufficient training data, and over-fitting, which are ineluctable for deep learning. Due to the co-existence of limited-angle problem and data truncation problem in this work, where truncation is caused by the large scale ice for immobilization of samples, applying iterative reconstruction such as simultaneous algebraic reconstruction technique with total variation regularization for data consistent reconstruction \cite{huang2019data} to improve such incorrect structures is not feasible.

In limited angle tomography, only structures whose orientations are tangent to available X-rays can be reconstructed \cite{quinto1993singularities,quinto2006introduction,quinto2007local,huang2016image}. Therefore, in the FBP reconstructions, most edges whose orientations are inside the scanned angular range are reconstructed. Because of this, for the chlorella reconstruction, several slices in the sagittal view contain good resolution structures. On the other hand, with the geometry setting in this work, the sagittal slices are equivalent to focus planes in tomosynthesis \cite{grant1972tomosynthesis} where most X-rays focus. Therefore, structures viewed in sagittal planes preserve better resolution than any horizontal planes. However, structures are preserved well only in a limited number of central slices in the sagittal view, while most structures are still distorted due to missing data (Fig.~\ref{Fig:tomosynsthesisSlice150}(a)). In order to view structures in any intersectional planes, artifact reduction is necessary. 

Due to missing data, many essential subcellular structures are distorted or even missing in the FBP reconstruction, e.\,g., the lipid bodies in this work. The distribution and states of subcellular structures provide crucial information of intracellular activities \cite{ortega2009bio,wang2015use}. With the power of deep learning in image processing, the proposed reconstruction method is competent for 3-D visualization of subcellular structures, as displayed in Fig.~\ref{Fig:paraView}. This observation indicates its important value for nano-scale imaging in biology, nanoscience and materials science.

\section{Conclusion And Outlook}

In this work, deep learning has been the first time to be applied to limited angle reconstruction in TXMs. PWLS preprocessing is beneficial to improve the image quality of deep learning reconstruction. Despite the limitation to accessing sufficient real training data, this work demonstrates that training a deep neural network model from synthetic data with proper noise modelling is a promising approach. The proposed deep learning reconstruction method remarkably improves the 3-D visualization of subcellular structures, indicating its important value for nano-scale imaging in biology, nanoscience and materials science.

Although promising and intriguing results are achieved in this work, the limited angle reconstruction problem is still not entirely resolved, since some structures are reconstructed inaccurately. In the future, the following aspects of work are worth investigating:
 
 \begin{itemize} 
 
 \item Evaluate the proposed deep learning reconstruction method on more complex samples is the next step.
 
 \item More realistic noise modelling for synthetic data should potentially improve deep learning performance.
 
 \item Explore new approaches to achieve data consistent reconstruction \cite{huang2019data} in the co-existence of limited-angle problem and data truncation problem.
 
 \item If possible, building up a database from complete real scans for training deep neural networks is necessary.
 \end{itemize}




\textbf{\ack{Acknowledgements}}

	We are very grateful for the chlorella data provided by the soft X-ray microscope at beamline BL07W in the National Synchrotron Radiation Laboratory (NSRL) in Hefei, China.
	
	The research leading to these results has received funding from the European Research Council (ERC) under the European Union’s Horizon 2020 research and innovation programme (ERC grant no. 810316).





\bibliographystyle{iucr}
\bibliography{hf5394}

@article{yu2018automatic,
  title={Automatic projection image registration for nanoscale X-ray tomographic reconstruction},
  author={Yu, Haiyan and Xia, Sihao and Wei, Chenxi and Mao, Yuwei and Larsson, Daniel and Xiao, Xianghui and Pianetta, Piero and Yu, Y-S and Liu, Yijin},
  journal={J. Synchrotron Rad.},
  volume={25},
  number={6},
  year={2018},
  publisher={International Union of Crystallography}
}

@article{yang2015registration,
  title={Registration of the rotation axis in X-ray tomography},
  author={Yang, Yimeng and Yang, Feifei and Hingerl, Ferdinand F and Xiao, Xianghui and Liu, Yijin and Wu, Ziyu and Benson, Sally M and Toney, Michael F and Andrews, Joy C and Pianetta, Piero},
  journal={J. Synchrotron Rad.},
  volume={22},
  number={2},
  pages={452--457},
  year={2015},
  publisher={International Union of Crystallography}
}

@article{quinto1993singularities,
  title={Singularities of the X-ray transform and limited data tomography in R\^{}2 and R\^{}3},
  author={Quinto, Eric Todd},
  journal={SIAM J. Math. Anal.},
  volume={24},
  number={5},
  pages={1215--1225},
  year={1993},
  publisher={SIAM}
}

@article{ioffe2015batch,
  title={Batch normalization: Accelerating deep network training by reducing internal covariate shift},
  author={Ioffe, Sergey and Szegedy, Christian},
  journal={arXiv preprint arXiv:1502.03167},
  year={2015}
}

@article{odena2016deconvolution,
  title={Deconvolution and checkerboard artifacts},
  author={Odena, Augustus and Dumoulin, Vincent and Olah, Chris},
  journal={Distill},
  volume={1},
  number={10},
  pages={e3},
  year={2016}
}

@article{ortega2009bio,
  title={Bio-metals imaging and speciation in cells using proton and synchrotron radiation X-ray microspectroscopy},
  author={Ortega, Richard and Deves, Guillaume and Carmona, Asunci{\'o}n},
  journal={J. Royal Soc. Interface},
  volume={6},
  number={suppl\_5},
  pages={S649--S658},
  year={2009},
  publisher={The Royal Society}
}

@article{wang2015use,
  title={Use of synchrotron radiation-analytical techniques to reveal chemical origin of silver-nanoparticle cytotoxicity},
  author={Wang, Liming and Zhang, Tianlu and Li, Panyun and Huang, Wanxia and Tang, Jinglong and Wang, Pengyang and Liu, Jing and Yuan, Qingxi and Bai, Ru and Li, Bai and others},
  journal={ACS nano},
  volume={9},
  number={6},
  pages={6532--6547},
  year={2015},
  publisher={ACS Publications}
}

@article{barnard1992360,
  title={A $360^\circ$ single-axis tilt stage for the high-voltage electron microscope},
  author={Barnard, David P and Turner, James N and Frank, Joachim and McEwen, Bruce F},
  journal={J. Micros.},
  volume={167},
  number={1},
  pages={39--48},
  year={1992},
  publisher={Wiley Online Library}
}

@inproceedings{hu2018squeeze,
  title={Squeeze-and-excitation networks},
  author={Hu, Jie and Shen, Li and Sun, Gang},
  booktitle={Proc CVPR},
  pages={7132--7141},
  year={2018}
}

@article{koster1997perspectives,
  title={Perspectives of molecular and cellular electron tomography},
  author={Koster, Abraham J and Grimm, Rudo and Typke, Dieter and Hegerl, Reiner and Stoschek, Arne and Walz, Jochen and Baumeister, Wolfgang},
  journal={J. Struct. Biol.},
  volume={120},
  number={3},
  pages={276--308},
  year={1997},
  publisher={Elsevier}
}

@article{holler2017omny,
  title={OMNY PIN—A versatile sample holder for tomographic measurements at room and cryogenic temperatures},
  author={Holler, Mirko and Raabe, J{\"o}rg and Wepf, Roger and Shahmoradian, Sarah H and Diaz, Ana and Sarafimov, Blagoj and Lachat, T and Walther, H and Vitins, M},
  journal={Rev. Sci. Instrum.},
  volume={88},
  number={11},
  pages={113701},
  year={2017},
  publisher={AIP Publishing}
}

@article{falk2019u,
  title={U-Net: deep learning for cell counting, detection, and morphometry},
  author={Falk, Thorsten and Mai, Dominic and Bensch, Robert and {\c{C}}i{\c{c}}ek, {\"O}zg{\"u}n and Abdulkadir, Ahmed and Marrakchi, Yassine and B{\"o}hm, Anton and Deubner, Jan and J{\"a}ckel, Zoe and Seiwald, Katharina and others},
  journal={Nat. methods},
  volume={16},
  number={1},
  pages={67},
  year={2019},
  publisher={Nature Publishing Group}
}

@article{shearing2011using,
  title={Using synchrotron X-ray nano-CT to characterize SOFC electrode microstructures in three-dimensions at operating temperature},
  author={Shearing, PR and Bradley, RS and Gelb, J and Lee, SN and Atkinson, A and Withers, PJ and Brandon, NP},
  journal={Electrochem. Solid-State Lett.},
  volume={14},
  number={10},
  pages={B117--B120},
  year={2011},
  publisher={The Electrochemical Society}
}

@article{brisard2012morphological,
  title={Morphological quantification of hierarchical geomaterials by X-ray nano-CT bridges the gap from nano to micro length scales},
  author={Brisard, Sebastien and Chae, Rosie S and Bihannic, Isabelle and Michot, Laurent and Guttmann, Peter and Thieme, J{\"u}rgen and Schneider, Gerd and Monteiro, Paulo JM and Levitz, Pierre},
  journal={Am. Mineral.},
  volume={97},
  number={2-3},
  pages={480--483},
  year={2012},
  publisher={Mineralogical Society of America}
}

@article{wang20153d,
  title={3D imaging of a rice pollen grain using transmission X-ray microscopy},
  author={Wang, Shengxiang and Wang, Dajiang and Wu, Qiao and Gao, Kun and Wang, Zhili and Wu, Ziyu},
  journal={J. Synchrotron Rad.},
  volume={22},
  number={4},
  pages={1091--1095},
  year={2015},
  publisher={International Union of Crystallography}
}

@article{shapiro2005biological,
  title={Biological imaging by soft x-ray diffraction microscopy},
  author={Shapiro, David and Thibault, Pierre and Beetz, Tobias and Elser, Veit and Howells, Malcolm and Jacobsen, Chris and Kirz, Janos and Lima, Enju and Miao, Huijie and Neiman, Aaron M and others},
  journal={Proc. Natl. Acad. Sci.},
  volume={102},
  number={43},
  pages={15343--15346},
  year={2005},
  publisher={National Acad Sciences}
}

@article{de2008nanoscale,
  title={Nanoscale chemical imaging of a working catalyst by scanning transmission X-ray microscopy},
  author={de Smit, Emiel and Swart, Ingmar and Creemer, J Fredrik and Hoveling, Gerard H and Gilles, Mary K and Tyliszczak, Tolek and Kooyman, Patricia J and Zandbergen, Henny W and Morin, Cynthia and Weckhuysen, Bert M and others},
  journal={Nature},
  volume={456},
  number={7219},
  pages={222},
  year={2008},
  publisher={Nature Publishing Group}
}

@article{sakdinawat2010nanoscale,
  title={Nanoscale X-ray imaging},
  author={Sakdinawat, Anne and Attwood, David},
  journal={Nature photonics},
  volume={4},
  number={12},
  pages={840},
  year={2010},
  publisher={Nature Publishing Group}
}

@article{andrews2011transmission,
  title={Transmission X-ray microscopy for full-field nano imaging of biomaterials},
  author={Andrews, Joy C and Meirer, Florian and Liu, Yijin and Mester, Zoltan and Pianetta, Piero},
  journal={Microsc. Res. Tech.},
  volume={74},
  number={7},
  pages={671--681},
  year={2011},
  publisher={Wiley Online Library}
}

@article{nelson2012operando,
  title={In operando X-ray diffraction and transmission X-ray microscopy of lithium sulfur batteries},
  author={Nelson, Johanna and Misra, Sumohan and Yang, Yuan and Jackson, Ariel and Liu, Yijin and Wang, Hailiang and Dai, Hongjie and Andrews, Joy C and Cui, Yi and Toney, Michael F},
  journal={J. Am. Chem. Soc.},
  volume={134},
  number={14},
  pages={6337--6343},
  year={2012},
  publisher={ACS Publications}
}

@article{wang2016nanotechnology,
  title={Nanotechnology: a new opportunity in plant sciences},
  author={Wang, Peng and Lombi, Enzo and Zhao, Fang-Jie and Kopittke, Peter M},
  journal={Trends Plant Sci.},
  volume={21},
  number={8},
  pages={699--712},
  year={2016},
  publisher={Elsevier}
}

@article{wang2000soft,
  title={Soft X-ray microscopy with a cryo scanning transmission X-ray microscope: II. Tomography},
  author={Wang, Y and Jacobsen, C and Maser, J and Osanna, A},
  journal={J. Microsc.},
  volume={197},
  number={1},
  pages={80--93},
  year={2000},
  publisher={BLACKWELL SCIENTIFIC}
}

@article{chao2005soft,
  title={Soft X-ray microscopy at a spatial resolution better than 15 nm},
  author={Chao, Weilun and Harteneck, Bruce D and Liddle, J Alexander and Anderson, Erik H and Attwood, David T},
  journal={Nature},
  volume={435},
  number={7046},
  pages={1210},
  year={2005},
  publisher={Nature Publishing Group}
}

@article{liu2018quantitative,
  title={Quantitative imaging of Candida utilis and its organelles by soft X-ray Nano-CT},
  author={Liu, J and Li, F and Chen, L and Guan, Y and Tian, L and Xiong, Y and Liu, G and Tian, Y},
  journal={J. Microsc.},
  volume={270},
  number={1},
  pages={64--70},
  year={2018},
  publisher={Wiley Online Library}
}

@article{grant1972tomosynthesis,
  title={Tomosynthesis: a three-dimensional radiographic imaging technique},
  author={Grant, David G},
  journal={IEEE Trans. Biomed. Eng.},
  number={1},
  pages={20--28},
  year={1972},
  publisher={IEEE}
}

@article{fei2006one,
  title={One-shot learning of object categories},
  author={Li, Fei-Fei and Fergus, Rob and Perona, Pietro},
  journal={IEEE Trans. Pattern Anal. Mach. Intell.},
  volume={28},
  number={4},
  pages={594--611},
  year={2006},
  publisher={IEEE}
}

@inproceedings{palatucci2009zero,
  title={Zero-shot learning with semantic output codes},
  author={Palatucci, Mark and Pomerleau, Dean and Hinton, Geoffrey E and Mitchell, Tom M},
  booktitle={Adv. Neural Inf. Process Syst.},
  pages={1410--1418},
  year={2009}
}

@article{pan2009survey,
  title={A survey on transfer learning},
  author={Pan, Sinno Jialin and Yang, Qiang},
  journal={ IEEE Trans. Knowl. Data Eng.},
  volume={22},
  number={10},
  pages={1345--1359},
  year={2009},
  publisher={IEEE}
}

@article{baudelet2017new,
  title={A new insight into cell walls of Chlorophyta},
  author={Baudelet, Paul-Hubert and Ricochon, Guillaume and Linder, Michel and Muniglia, Lionel},
  journal={Algal Res.},
  volume={25},
  pages={333--371},
  year={2017},
  publisher={Elsevier}
}

@article{maier2019learning,
  title={Learning with Known Operators reduces Maximum Training Error Bounds},
  author={Maier, Andreas K and Syben, Christopher and Stimpel, Bernhard and W{\"u}rfl, Tobias and Hoffmann, Mathis and Schebesch, Frank and Fu, Weilin and Mill, Leonid and Kling, Lasse and Christiansen, Silke},
  journal={Nat. Mach. Intell.},
  year={2019}
}

@article{wang2006penalized,
  title={Penalized weighted least-squares approach to sinogram noise reduction and image reconstruction for low-dose X-ray computed tomography},
  author={Wang, Jing and Li, Tianfang and Lu, Hongbing and Liang, Zhengrong},
  journal={IEEE Trans. Med. Imaging},
  volume={25},
  number={10},
  pages={1272--1283},
  year={2006},
  publisher={IEEE}
}

@article{wang2019jitter,
  title={Jitter correction for transmission X-ray microscopy via measurement of geometric moments},
  author={Wang, Shengxiang and Liu, Jianhong and Li, Yinghao and Chen, Jian and Guan, Yong and Zhu, Lei},
  journal={J. Synchrotron Rad.},
  volume={26},
  number={5},
  year={2019},
  publisher={International Union of Crystallography}
}

@article{papoulis1975new,
  title={A new algorithm in spectral analysis and band-limited extrapolation},
  author={Papoulis, Athanasios},
  journal={IEEE Trans. Circuits Syst.},
  volume={22},
  number={9},
  pages={735--742},
  year={1975},
  publisher={IEEE}
}

@article{wurfl2018deep,
  title={Deep learning computed tomography: Learning projection-domain weights from image domain in limited angle problems},
  author={W{\"u}rfl, Tobias and Hoffmann, Mathis and Christlein, Vincent and Breininger, Katharina and Huang, Yixin and Unberath, Mathias and Maier, Andreas K},
  journal={IEEE Trans. Med. Imaging},
  volume={37},
  number={6},
  pages={1454--1463},
  year={2018},
  publisher={IEEE}
}

@article{gerchberg1974super,
  title={Super-resolution through error energy reduction},
  author={Gerchberg, RW},
  journal={J. Mod. Opt.},
  volume={21},
  number={9},
  pages={709--720},
  year={1974},
  publisher={Taylor \& Francis}
}

@article{defrise1983regularized,
  title={A regularized iterative algorithm for limited-angle inverse Radon transform},
  author={Defrise, Michel and De Mol, Christine},
  journal={Opt. Acta: Int. J. Opt.},
  volume={30},
  number={4},
  pages={403--408},
  year={1983},
  publisher={Taylor \& Francis}
}

@article{kudo1991sinogram,
  title={Sinogram recovery with the method of convex projections for limited-data reconstruction in computed tomography},
  author={Kudo, Hiroyuki and Saito, Tsuneo},
  journal={J. Opt. Soc. Am. A Opt. Image Sci. Vis.},
  volume={8},
  number={7},
  pages={1148--1160},
  year={1991},
  publisher={Optical Society of America}
}

@article{huang2019traditional,
  title={Traditional machine learning for limited angle tomography},
  author={Huang, Yixing and Lu, Yanye and Taubmann, Oliver and Lauritsch, Guenter and Maier, Andreas},
  journal={Int. J. Comput. Assist. Radiol. Surg.},
  volume={14},
  number={1},
  pages={11--19},
  year={2019},
  publisher={Springer}
}

@inproceedings{Huang2018Papoulis,
	pages={189--192},
	booktitle ={Proc. CT Meeting},
	year={2018},
	location={Salt Lake City, Utah, the USA},
	title={{Papoulis-Gerchberg Algorithms for Limited Angle Tomography Using Data Consistency Conditions}},
	author={Yixing Huang and Oliver Taubmann and Xiaolin Huang and Guenter Lauritsch and Andreas Maier}
}

@article{huang2018scale,
  title={Scale-space anisotropic total variation for limited angle tomography},
  author={Huang, Yixing and Taubmann, Oliver and Huang, Xiaolin and Haase, Viktor and Lauritsch, Guenter and Maier, Andreas},
  journal={IEEE Trans. Radiat. Plasma Med. Sci.},
  volume={2},
  number={4},
  pages={307--314},
  year={2018},
  publisher={IEEE}
}

@article{bubba2019learning,
  title={Learning the invisible: A hybrid deep learning-shearlet framework for limited angle computed tomography},
  author={Bubba, Tatiana A and Kutyniok, Gitta and Lassas, Matti and M{\"a}rz, Maximilian and Samek, Wojciech and Siltanen, Samuli and Srinivasan, Vignesh},
  journal={Inverse Probl.},
  volume={35},
  number={6},
  pages={064002},
  year={2019},
  publisher={IOP Publishing}
}

@article{huang2019data,
  title={Data Consistent Artifact Reduction for Limited Angle Tomography with Deep Learning Prior},
  author={Huang, Yixing and Preuhs, Alexander and Lauritsch, Guenter and Manhart, Michael and Huang, Xiaolin and Maier, Andreas},
  journal={arXiv preprint arXiv:1908.06792},
  year={2019}
}

@inproceedings{huang2018some,
  title={Some investigations on robustness of deep learning in limited angle tomography},
  author={Huang, Yixing and W{\"u}rfl, Tobias and Breininger, Katharina and Liu, Ling and Lauritsch, G{\"u}nter and Maier, Andreas},
  booktitle={Proc. MICCAI},
  pages={145--153},
  year={2018},
  publisher = {Springer International Publishing},
  address ={Cham}
}

@article{Qu2008An,
  title={An iterative algorithm for angle-limited three-dimensional image reconstruction},
  author={Qu, Gang Rong and Lan, Yong Sheng and Jiang, Ming},
  journal={Acta Math. Appl. Sin.},
  volume={24},
  number={1},
  pages={157-166},
  year={2008},
}

@article{Qu2009Landweber,
  title={Landweber Iterative Methods for Angle-limited Image Reconstruction},
  author={Qu, Gang Rong and Jiang, Ming},
  journal={Acta Math. Appl. Sin.},
  volume={25},
  number={2},
  pages={327-334},
  year={2009},
}

@article{huang2017Restoration,
  author={Yixing Huang and Xiaolin Huang and Oliver Taubmann and Yan Xia and Viktor Haase and Joachim Hornegger and Guenter
Lauritsch and Andreas Maier},
  title={Restoration of missing data in limited angle tomography based on Helgason–Ludwig consistency conditions},
  journal={Biomed. Phys. $\&$ Eng. Express},
  volume={3},
  number={3},
  pages={035015},
  year={2017},
}

@article{louis1980picture,
  title={Picture reconstruction from projections in restricted range},
  author={Louis, Alfred K and T{\"o}rnig, W},
  journal={Math. Methods Appl. Sci.},
  volume={2},
  number={2},
  pages={209--220},
  year={1980},
  publisher={Wiley Online Library}
}

@article{willsky1990constrained,
  title={Constrained sinogram restoration for limited-angle tomography},
  author={Willsky, Alan S and Prince, Jerry L},
  journal={Opt. Eng.},
  volume={29},
  number={5},
  pages={535--544},
  year={1990},
  publisher={International Society for Optics and Photonics}
}

@inproceedings{tobias2016deep,
	author={Tobias W{\"u}rfl and Florin Cristian Ghesu and Vincent Christlein and Andreas Maier},
	editor={Springer},
	location={Athen},
	volume={3},
	booktitle={Proc. MICCAI},
	title={{Deep Learning Computed Tomography}},
	pages={432--440},
	year={2016},
}

@inproceedings{ronneberger2015u,
  title={{U-Net}: Convolutional networks for biomedical image segmentation},
  author={Ronneberger, Olaf and Fischer, Philipp and Brox, Thomas},
  booktitle={Proc. MICCAI},
  pages={234--241},
  year={2015},
  organization={Springer}
}

@inproceedings{gu2017multi,
  title={Multi-Scale Wavelet Domain Residual Learning for Limited-Angle CT Reconstruction},
  author={Gu, Jawook and Ye, Jong Chul},
  booktitle={Proc. Fully3D},
  location = {Xi'an, China},
  year={2017},
}

@incollection{louis1981approximation,
  title={Approximation of the Radon transform from samples in limited range},
  author={Louis, Alfred K},
  booktitle={Mathematical Aspects of Computerized Tomography},
  pages={127--139},
  year={1981},
  publisher={Springer}
}

@article{davison1983ill,
  title={The ill-conditioned nature of the limited angle tomography problem},
  author={Davison, Mark E},
  journal={SIAM J. Appl. Math.},
  volume={43},
  number={2},
  pages={428--448},
  year={1983},
  publisher={SIAM}
}

@article{louis1986incomplete,
  title={Incomplete data problems in {X-ray} computerized tomography: I. Singular Value Decomposition of the Limited Angle Transform},
  author={Louis, Alfred K},
  journal={Numerische Mathematik},
  volume={48},
  number={3},
  pages={251--262},
  year={1986},
  publisher={Springer}
}

@book{natterer1986mathematics,
  title={The mathematics of computerized tomography},
  author={Natterer, Frank},
  volume={32},
  year={1986},
  publisher={SIAM}
}

@article{sidky2006accurate,
  title={Accurate image reconstruction from few-views and limited-angle data in divergent-beam {CT}},
  author={Sidky, Emil Y and Kao, Chien-Min and Pan, Xiaochuan},
  journal={J. Xray Sci. Technol.},
  volume={14},
  number={2},
  pages={119--139},
  year={2006},
  publisher={IOS Press}
}

@incollection{huang2016image,
  title={Image Quality Analysis of Limited Angle Tomography Using the Shift-Variant Data Loss Model},
  author={Huang, Yixing and Lauritsch, Guenter and Amrehn, Mario and Taubmann, Oliver and Haase, Viktor and Stromer, Daniel and Huang, Xiaolin and Maier, Andreas},
  booktitle={Proc. BVM},
  pages={277--282},
  year={2016},
  publisher={Springer}
}

@article{chen2013limited,
  title={A limited-angle {CT} reconstruction method based on anisotropic {TV} minimization},
  author={Chen, Zhiqiang and Jin, Xin and Li, Liang and Wang, Ge},
  journal={Phys. Med. Biol.},
  volume={58},
  number={7},
  pages={2119},
  year={2013},
  publisher={IOP Publishing}
}

@inproceedings{huang2016watv,
	author={Huang, Yixing and Taubmann, Oliver and Huang, Xiaolin and Haase, Viktor and Lauritsch, Guenter and Maier, Andreas},
	location={Prague},
	booktitle={Proc. ISBI},
	title={{A new weighted anisotropic total variation algorithm for limited angle tomography}},
	pages={585--588},
	year={2016}
}

@article{ritschl2011improved,
  title={Improved total variation-based {CT} image reconstruction applied to clinical data},
  author={Ritschl, Ludwig and Bergner, Frank and Fleischmann, Christof and Kachelrie{\ss}, Marc},
  journal={Phys. Med. Biol.},
  volume={56},
  number={6},
  pages={1545},
  year={2011},
  publisher={IOP Publishing}
}

@article{sidky2008image,
  title={Image reconstruction in circular cone-beam computed tomography by constrained, total-variation minimization},
  author={Sidky, Emil Y and Pan, Xiaochuan},
  journal={Phys. Med. Biol.},
  volume={53},
  number={17},
  pages={4777},
  year={2008},
  publisher={IOP Publishing}
}

@inproceedings{quinto2006introduction,
  title={An introduction to {X-ray} tomography and Radon transforms},
  author={Quinto, Eric Todd},
  booktitle={Proc. symp. Appl. Math.},
  volume={63},
  pages={1},
  year={2006}
}

@article{quinto2007local,
  title={Local algorithms in exterior tomography},
  author={Quinto, Eric Todd},
  journal={J. Comput. Appl. Math.},
  volume={199},
  number={1},
  pages={141--148},
  year={2007},
  publisher={Elsevier}
}

@article{helgason1965radon,
  title={The Radon transform on Euclidean spaces, compact two-point homogeneous spaces and Grassmann manifolds},
  author={Helgason, Sigurdur},
  journal={Acta Math.},
  volume={113},
  number={1},
  pages={153--180},
  year={1965},
  publisher={Springer}
}

@article{ludwig1966radon,
  title={The Radon transform on Euclidean space},
  author={Ludwig, Donald},
  journal={Comm. Pure Appl. Math.},
  volume={19},
  number={1},
  pages={49--81},
  year={1966},
  publisher={Wiley Online Library}
}

\end{document}